\documentclass[a4paper,fleqn,usenatbib]{mnras}
\usepackage[T1]{fontenc}
\usepackage{ae,aecompl}
\usepackage{graphicx}
\usepackage{amsmath}
\usepackage{amssymb}
\usepackage{txfonts}
\usepackage{float}
\usepackage{adjustbox}
\usepackage{siunitx}
\usepackage{tensor}

\usepackage{tensor}
\usepackage{nicefrac}

\graphicspath{{figures/}}

\usepackage{import}  
\usepackage{pdfpages}

\newcommand{\ltsima}{$\; \buildrel < \over \sim \;$}
\newcommand{\lsim}{\lower.5ex\hbox{\ltsima}}
\newcommand{\gtsima}{$\; \buildrel > \over \sim \;$}
\newcommand{\gsim}{\lower.5ex\hbox{\gtsima}}
\newcommand{\bra}{\langle}
\newcommand{\ket}{\rangle}

\newcommand{\dd}{\mathrm{d}}

\newcommand{\likeli}{\mathcal{L}}
\newcommand{\w}{w}

\newcommand{\im}{\mathrm{im}\:}
\DeclareMathOperator{\cosi}{Ci}

\title[partitions for functional inference]
{Partition function approach to non-Gaussian likelihoods: partitions for the inference of functions and the Fisher-functional}
\author[R.M.  Kuntz, M.Ph. Herzog, H. v. Campe, L. R{\"o}ver, B.M. Sch{\"a}fer]
{Rebecca Maria Kuntz$^2$\thanks{e-mail: kuntz@stud.uni-heidelberg.de}, Maximilian Philipp Herzog$^2$, Heinrich von Campe$^2$,\newauthor Lennart R{\"o}ver$^{1,2}$, Bj{\"o}rn Malte Sch{\"a}fer$^2$\thanks{e-mail: bjoern.malte.schaefer@uni-heidelberg.de}\\
$^1$ Institut f{\"u}r Theoretische Physik, Universit{\"a}t Heidelberg, Philosophenweg 16, 69120 Heidelberg, Germany\\
$^2$Zentrum f{\"u}r Astronomie der Universit{\"a}t Heidelberg, Astronomisches Rechen-Institut, Philosophenweg 12, 69120 Heidelberg, Germany}

\begin{document}
\pagerange{\pageref{firstpage}--\pageref{lastpage}}
\pubyear{2022}
\maketitle
\label{firstpage}

\begin{abstract}
Motivated by constraints on the dark energy equation of state from a data set of supernova distance moduli, we propose a formalism for the Bayesian inference of functions: Starting at a functional variant of the Kullback-Leibler divergence we construct a functional Fisher-matrix and a suitable partition functional which takes on the shape of a path integral. After showing the validity of the Cram{\'e}r-Rao bound and unbiasedness for functional inference in the Gaussian case, we construct Fisher-functionals for the dark energy equation of state constrained by the cosmological redshift-luminosity relationship of supernovae of type Ia, for both the linearised and the lowest-order non-linear model. Introducing Fourier-expansions and expansions into Gegenbauer-polynomials as discretisations of the dark energy equation of state function shows how the uncertainty on the inferred function scales with model complexity and how functional assumptions can lead to errors in extrapolation to poorly constrained redshift ranges.
\end{abstract}


\onecolumn

\section{introduction}
Inference problems are central to cosmology, where one often faces problems related to highly-dimensional, degenerate parameter spaces and non-Gaussian likelihoods, and cases where the influence of cosmological and astrophysical parameters can not be easily disentangled. Of particular relevance to cosmology are questions in relation to dark energy and its properties, which can be investigated in the dynamics of the background Friedmann-Lema{\^i}tre-Robertson-Walker (FLRW) spacetime. The homogeneous and isotropic expansion dynamics of the background leave an imprint either directly in the distance-redshift relation of supernovae of type Ia or indirectly through their influence on cosmic structure formation \citep{linder_cosmic_2005, barnes_influence_2005}. Dark energy summarises all effects of repulsive gravity on cosmological scales $\sim \chi_H = c/H_0$ and is often modelled in terms of an ideal, cosmological fluid or a scalar, self-interacting field, both exhibiting homogeneity and isotropy as the fundamental symmetries of FLRW-spacetimes.

The source of gravity in general relativity are ideal fluids, characterised with a divergence-free energy-momentum tensor $T_{\mu\nu}$, with $g^{\alpha\mu}\nabla_\alpha T_{\mu\nu} = 0$, that contains only the velocity, the density, and the pressure of the fluid. For a FLRW-fluid in the comoving frame, the spatial velocities vanish and the only remaining degrees of freedom are density $\rho$ and pressure $p$, which both can only depend on time $t$, as a consequence of the cosmological principle. Often, the density of the fluid is re-expressed in terms of the critical density $\rho_\mathrm{crit} = 3H(a)^2/(8\pi G)$ with the Hubble-function $H(a)$, to yield the definition of the density parameter $\w = \rho/\rho_\mathrm{crit}$. The ratio between pressure and density defines the equation of state parameter $\w = p/(\rho c^2)$. For critical universes that are filled up to the critical density $\rho = \rho_\mathrm{crit}$, or equivalently, $\Omega = 1$ with a single fluid whose equation of state parameter is constant in time and reflects an internal property of the fluid, one can derive that $3(1+\w) = 2(1+q)$ with the deceleration parameter $q = -\ddot{a}a/\dot{a}^2$ from the scale factor $a(t)$ and its derivatives. One clearly sees that $\w = -1/3$ is a peculiar value that separates decelerated universes with $\ddot{a} > 0$ for $\w > -1/3$ from accelerated universes with $\ddot{a} < 0$ for $\w < -1/3$: Effects of repulsive gravity are associated with equations of state more negative than $\w = -1/3$.

In many intuitive cases, the equation of state parameter reflects a typical and immutable property of the fluid, for instance $\w = +1/3$ for radiation, $\w = 0$ for non-relativistic matter, or $\w = -1$ for the cosmological constant $\Lambda$, but from the point of general relativity, this restriction is not necessary: Any time-evolving equation of state is permissible which has led to a number of parameterisations, the most popular being proposed by \citet{chevallier_accelerating_2001, linder_exploring_2003}, who assume linear changes of $\w$ with scale factor $a$,
\begin{equation}
\w(a) = \w_0 + (1- a)\w_a,
\label{eqn_cpl}
\end{equation}
such that the choice of $\w_0 = -1$ and $\w_a = 0$ recovers the $\Lambda$-case: Clearly, measurements that constrain $\w_0$ to be close to $-1$ and $\w_a$ to be zero make a strong case for a non-time evolving dark energy component to be identified with the cosmological constant.

Transitioning from a fluid as the source of gravity to a (homogeneous) scalar field $\varphi$ with self-interaction $V(\varphi)$ as the source of gravity was opened by quintessence constructions \citep{WetterichQuintessence, PeeblesQuintessence, linder_dynamics_2008, tsujikawa_quintessence:_2013, mortonson_dark_2013}: If coupled to gravity, the field sources gravity like a fluid with an equation of state parameter
\begin{equation}
\w = \frac{\dot{\varphi}^2 - 2V(\varphi)}{\dot{\varphi}^2 + 2V(\varphi)},
\end{equation}
and if the Lagrange-density does not explicitly depend on the coordinates, the energy-momentum tensor associated with the scalar $\varphi$ is covariantly conserved, $g^{\alpha\mu}\nabla_\alpha T_{\mu\nu} = 0$. As the field evolves according to the Klein-Gordon-type equation $\ddot{\varphi} + 3H(t)\dot{\varphi} = -\dd V/\dd\varphi$, it generates a time-evolving $\w$ dependent on the initial conditions and the shape of the interaction potential. The case of a cosmological constant $\Lambda$ is recovered in the slow-roll limit $\dot{\varphi}^2\ll V(\varphi)$.

Therefore, varying equation of state parameters $\w$ are natural in both the fluid case and the field case. General relativity does not restrict $\w$ to obey a particular parameterisation and, as far as gravity is concerned, $\w$ can be a free function of time; with only two possible conditions: $(i)$ $\w$ should not be more negative than $-1$, otherwise the scale factor $a(t)$ diverges after a finite time, which is referred to as phantom dark energy, and $(ii)$ values of $\w$ more positive than $+1/3$ would lead to a negative Ricci-scalar $R$. In consequence, the cases $\w = 1/3$ with $q = 1$ and $\w=-1$ with $q = -1$ are limits of the otherwise free function $\w(a)$.

Choices for the parameterisation of $\w(a)$ (or related reconstructions of the Hubble-function $H(a)$ or derived quantities like distance measures or the growth function $D_+(a)$) that have been used in the literature include higher-order Taylor expansions generalising eqn.~(\ref{eqn_cpl}), Fourier-expansions or orthogonal polynomials \citep{2002Gerke, corasaniti_model-independent_2003, melchiorri_state_2003,  amendola_fitting_2004, bean_probing_2004, pogosian_tracking_2005, liddle_present_2006, mignone_model-independent_2008, Maturi_2009, biesiada_cosmic_2010, lee_optimal_2011, Benitez_Herrera_2011, Benitez_Herrera_2013, Li_2014,mukherjee_reconstruction_2015, mukherjee_acceleration_2016, Feng_2016, hee_constraining_2016, matilla_geometry_2017, haude2019modelindependent, schmidt2020sensitivity}, or Gaussian processes \citep{Shafieloo_2006, holsclaw_nonparametric_2011, shafieloo_gaussian_2012} with a fixed assumption about the covariance. But in all cases, the assumption of a parameterisation is a restriction on the functional variability of $\w$ unvouched by relativity, and results do depend on the chosen parameterisation \citep{linder_biased_2006}. In extreme cases like early dark energy \citep{doran_early_2006}, their phenomenology is not captured by eqn.~(\ref{eqn_cpl}). This point is exactly the motivation of this paper: We would like to find out how one would formally infer the equation of state as a function from cosmological data, and how one would propagate the errors of a cosmological measurement to an uncertainty, or error-tube, around the inferred function, and how the inferred equation of state function and its error depends on a chosen parameterisation. For that purpose, we aim for a generalisation of the Fisher-matrix formalism and of an associated partition sum as an embodiment of the Bayes-theorem for statistical inference, in terms of functional derivatives and functional (path) integration. We would like to point out that functional derivatives have already found applications in cosmology, for instance in \citet{2023arXiv230104085R}, who defines a functional (or rather, distributional) Fisher-matrix for variations in the source redshift distribution in a weak lensing survey, including a mechanism for maintaining the normalisation of the distributions in variation. In fact, from this point of view we would like to think of a parameterisation as a choice of discretisation of the function to be inferred. After introducing the relevant concepts in Sect.~\ref{sect_infogeo}, we apply our formalism to supernova data \citep[for a review of the method, see][]{DavisParkinson2016} in Sect.~\ref{sect_supernova}, before summarising our main results in Sect.~\ref{sect_summary}.

Throughout the paper, we adopt the summation convention and make the choice to denote parameter tuples $\theta^\alpha$, with contravariant Greek indices; and data tuples $y^i$ with contravariant Latin indices. With these conventions, the Fisher-matrix $F_{\alpha\beta}$ is a covariant tensor, allowing to write quadratic forms like $\Delta\chi^2 = F_{\alpha\beta}\delta\theta^\alpha\delta\theta^\beta$, and to define an inverse $F^{\alpha\beta}$, $F_{\alpha\beta}F^{\beta\gamma} = \delta_\alpha^\gamma$. Likewise, the data covariance matrix is given by $C^{ij} = \bra y^iy^j\ket - \bra y^i\ket\bra y^j\ket$ with its inverse $C_{ij}$ defined by $C_{ij}C^{jk}=\delta_i^k$. As a reference cosmological model we use a spatially flat $\Lambda$CDM-cosmology with a fixed equation of state parameter $\w = -1$. As our primary interest is the variability of the equation of state function $\w(a)$, we work with a fixed value of $\Omega_m=0.3$ for the matter density, commensurate with current estimates, such that the prior on spatial flatness implies $\Omega_\mathrm{DE} = 1-\Omega_m = 0.7$. We choose to work with the scale factor $a$ as the variable denoting cosmic evolution because of its compact range of values, but we would like to emphasise that analogous formulas apply to the choice of redshift $z$ or cosmic time $t$.

\section{Canonical MCMC and its connection to information geometry}\label{sect_infogeo}

\subsection{From Bayes-theorem to the canonical partition function}
Bayes' theorem \citep[with relevance to cosmology, see][]{trotta_applications_2007, trotta_bayes_2008, trotta_bayesian_2017} relates the posterior distribution $p(\theta| y)$ for the parameters $\theta$ of a physical model given the observation of a data set $y$ to the likelihood $\mathcal{L}(y|\theta)$ for observing the data set for a given choice of parameters,
\begin{equation}
p(\theta| y) = \frac{\mathcal{L}(y|\theta)\pi(\theta)}{p(y)}
\end{equation}
by normalisation with the Bayesian evidence $p(y)$,
\begin{equation}
p(y) = \int\dd^n\theta\:\mathcal{L}(y|\theta)\pi(\theta)
\end{equation}
which corresponds to the probability of having observed the data set in the first place, under a given prior distribution $\pi(\theta)$. The particular mathematical structure of Bayes' law with an integral in the denominator and the integrand in the numerator suggests the definition of the partition function
\begin{equation}
Z[\beta,J_\alpha] = 
\int\dd^n\theta\:\left[\mathcal{L}(y|\theta)\pi(\theta)\:\exp\left(J_\alpha\theta^\alpha\right)\right]^\beta
\end{equation}
as a generalisation of the Bayesian evidence, which is recovered through $Z[\beta=1,J_\alpha=0] = p(y)$ for vanishing sources $J_\alpha$ and at unit (inverse) temperature $\beta = 1$. Then, statistical properties of the posterior distribution follow by differentiation of the Helmholtz-energy $F(\beta,J_\alpha) = -\ln Z[\beta,J_\alpha]/\beta$: Derivatives with respect to $J_\alpha$ yield the cumulants of the posterior distribution $p(\theta| y)$; derivatives with respect to $\beta$ the Shannon-entropy of $p(\theta| y)$, where both derivatives are taken again at unit (inverse) temperature $\beta = 1$ and for vanishing sources $J_\alpha  = 0$.

Interestingly, the partition function approach relates geometry with statistics: In fact, the Kullback-Leibler divergence $\Delta S$ \citep{baez_bayesian_2014} between two likelihoods differing by $\delta\theta$ in their parameter choices can be expanded in lowest order to be quadratic in $\delta\theta$ \citep[see][with applications to cosmology]{amari_information_2016, carron_information_2012, giesel_information_2021},
\begin{equation}
    \Delta S = 
    \int \mathrm{d}y\: \mathcal{L}(y|\theta) \: \ln \bigg(\frac{\mathcal{L}(y|\theta)}{\mathcal{L}(y|\theta+\delta \theta)}\bigg) \cong \frac{1}{2} F_{\mu\nu}(\theta)\:\delta\theta^\mu \delta \theta^\nu,
\label{eqn_kl_parameters}
\end{equation}
yielding the Fisher-metric $F_{\mu\nu}(\theta)$ as a positive definite, symmetric bilinear form, which introduces a notion of distance on the statistical manifold formed by the distributions $\mathcal{L}(y|\theta)$. If a global choice of coordinates is possible where $F_{\mu\nu}$ becomes constant, the metric distance $\Delta S$ corresponds to $\Delta\chi^2/2$, which assumes the form $\Delta\chi^2 = F_{\mu\nu}\delta\theta^\mu\delta\theta^\nu$ with $\delta\theta^\mu$ being the coordinate distance to the fiducial model values. The Fisher-metric itself is computable through \citep{tegmark_karhunen-loeve_1997, raveri_cosmicfish_2016, coe_fisher_2009, bassett_fisher_2011}
\begin{equation}
F_{\mu\nu} = 
\int\dd y\:\mathcal{L}(y|\theta)\:\frac{\partial\ln\mathcal{L}(y|\theta)}{\partial\theta^\mu}\frac{\partial\ln\mathcal{L}(y|\theta)}{\partial\theta^\nu}.
\label{eqn_fisher_matrix}
\end{equation}
A central and very interesting aspect of the partition function formalism are its analytical properties in the case of a Gaussian likelihood $\mathcal{L}(y|\theta)\propto\exp(-\chi^2(y|\theta)/2)$. For an equally Gaussian or uniform prior distribution $\pi(\theta)$ one find a closed form of the partition function by solving the Gaussian integral. Then, the logarithmic partition function is a second-order polynomial and the cumulant expansion truncates at second order: The posterior distribution is necessarily Gaussian for linear models.

The central idea of this paper is the replacement of a finite number of parameters $\theta^\mu$ by a parameter function $\w(a)$ on which the model depends. To that purpose, we need a generalisation of $(i)$ the Fisher-metric which we call the Fisher-functional, $(ii)$ a corresponding quadratic form for $\Delta\chi^2$ and $(iii)$ a suitably constructed functional canonical partition sum, which becomes effectively a path integral over the space of functions $\w(a)$. With these concepts in place, one can reintroduce a discretisation of $\w(a)$ in terms of Fourier-modes or a suitably constructed set of orthogonal polynomials.

\subsection{Functional inference}

\subsubsection{The Fisher-functional}
\label{subsubsection:fisher-functional}
The transition from discrete indices $1\leq\mu\leq n$ to continuous indices $a \in K$ from some compact index domain $K$ (in our case, $K = [0,1]$) corresponds to going from parameter-tuples $\theta \in \mathbb{R}^n$ to scalar, square-integrable parameter functions $\w \in L^2(K)$. In this picture, a likelihood $\mathcal{L}: L^2(K) \to P(\Omega),\; w \mapsto \likeli(y | w)$ maps those functions to probability densities on data space, effectively making the likelihood a distribution conditional on a function instead of a parameter. Square-integrability is chosen by convenience, as many of the physically relevant choices are Fourier-modes or orthonormal polynomials. As a consequence, one would expect the Fisher-metric to be defined as a functional over the space of functions $\delta\w(a)$, relative to a fiducial function $\w(a)$.

To derive the Fisher-functional $F[\w(a),\w(a')]$, we start from the functional analogue of the Kullback-Leibler divergence $\Delta S$ in eqn.~\eqref{eqn_kl_parameters} and expand it in lowest order to be quadratic in $\delta\w$, 
\begin{equation}
    \Delta S = 
    \int \mathrm{d}y\: \mathcal{L} (y|\w) \; \ln\bigg(\frac{\mathcal{L}(y|\w)}{\mathcal{L}(y|\w + \delta \w)}\bigg).
\end{equation}
The Einstein summation convention, which was implicit in the case of discrete indexing, is now made explicit  by integrals over the index domain $K$. One finds at order $\mathcal{O}\Big[(\delta\w)^3\Big]$
\begin{align}
    \Delta S \cong 
    \int \mathrm{d}y \: \mathcal{L}(y|\w) \:
    &\left[
    \int \mathrm{d}a\: \frac{\delta}{\delta(\w(a) +\delta\w(a))}
    \ln\bigg(\frac{\mathcal{L}(y|\w)}{\mathcal{L}(y|\w+\delta \w)}\bigg)\Bigg|_{\delta \w = 0}  \delta \w(a)\right. \nonumber\\
    &\;  +  
    \left. \frac{1}{2}\int \mathrm{d}a \int \mathrm{d}a'\: \frac{\delta^2}{\delta (\w(a)+\delta \w(a))\:\delta (\w(a')+\delta\w(a'))} \ln\bigg(\frac{\mathcal{L}(y|\w)}{\mathcal{L}(y|\w+\delta \w)}\bigg)\Bigg|_{\delta \w = 0}  \delta\w(a) \:\delta \w(a')\right].
\end{align}
When the functional derivatives are performed, all terms higher than second order in the perturbations are discarded just as in the case of partial differentiation with respect to discrete parameters. Swapping differentiation and integration, one reaches
\begin{align}
     \Delta S &\cong - \int \mathrm{d}a\: \delta \w(a) \;\frac{\delta}{\delta \w(a)} \bigg(\int\mathrm{d}y\:  \mathcal{L}(y|\w)\bigg) + \frac{1}{2} \int\mathrm{d}a\int \mathrm{d}a' \: \int\mathrm{d}y \:\mathcal{L}(y|\w) \:\frac{\delta \ln \mathcal{L}(y|\w)}{\delta \w(a)} \:\frac{\delta \ln\mathcal{L}(y|\w)}{\delta \w(a')} \delta \w(a)\: \delta \w(a')\nonumber\\
    &\hspace{0.3cm}- \frac{1}{2} \int \mathrm{d}a \int \mathrm{d}a' \:\delta \w(a) \:\delta \w(a')\: \frac{\delta^2}{\delta\w(a) \:\delta \w(a')} \bigg(\int \mathrm{d}y\:\mathcal{L}(y|\w)\bigg).
\end{align}
The likelihood $\mathcal{L}(y|\w)$ is normalized as a distribution over the data space, $\int \mathrm{d}y \: \mathcal{L}(y|\w) = 1$, leading to
\begin{equation}
    \Delta S \cong \int\mathrm{d}a \int \mathrm{d}a' \: \frac{1}{2}\int\mathrm{d}y \:\mathcal{L}(y|\w)  \:\frac{\delta \ln \mathcal{L}(y|\w)}{\delta \w(a)} \frac{\delta \ln \mathcal{L}(y|\w)}{\delta \w(a')}  \delta \w(a)\:\delta \w(a') \equiv \frac{1}{2} \int \mathrm{d}a  \int \mathrm{d}a'\:  F[\w(a),\w(a')]  \: \delta \w(a)\: \delta \w(a'),
\label{eqn_kl_functional}
\end{equation}
where the Fisher-functional can be identified as
\begin{equation}
    F [\w(a),\w(a')] = \int\mathrm{d}y \:\mathcal{L}(y|\w)\: \frac{\delta \ln\mathcal{L}(y|\w)}{\delta \w(a)} \frac{\delta \ln \mathcal{L}(y|\w)}{\delta \w(a')}.
\label{eqn_fisher_functional}
\end{equation}
The structural similarity of eqns.~(\ref{eqn_kl_functional}) and~(\ref{eqn_fisher_functional}) with eqns.~(\ref{eqn_kl_parameters}) and~(\ref{eqn_fisher_matrix}) is striking. The obtained Fisher functional is symmetric, $F[\w(a), \w(a')] = F[\w(a'), \w(a)]$, and positive semi-definite. Hence, for two functions $\eta, \rho \in L^2(K)$ we may define a scalar product \mbox{$\langle \eta,\rho \rangle_\w = \int \dd a \int \dd a'\: F[\w(a), \w(a')]\:\eta(a) \: \rho(a')$}, such that $L^2(K)$ becomes a Hilbert space. For details, we refer to Appendix~\ref{appendix_functional_fisher}.

Note that in general the Fisher-functional depends on the choice of reference function $\w$ relative to which the expressions are expanded; in our application the natural choice is $\w(a) \equiv -1$ corresponding to the $\Lambda$-case. In fact, this corresponds exactly to the choice of a fiducial cosmology typical for Fisher-type analyses, relative to which variations $\delta\w(a)$ are considered. In this way, each point has a local neighbourhood that is homeomorphic to $L^2(K)$, rather than $\mathbb{R}^n$, making $L^2(K)$ a manifold with geometric properties in the spirit of information geometry \citep{amari_information_2016}. If the likelihood is only evaluated at a finite number of data points $N$, the Fisher-functional will be degenerate and geometric objects cannot be contracted, preventing the usual construction of objects like the Riemann curvature tensor \citep{Stoica_2014}. 

\begin{figure}
    \centering
    \includegraphics[width =\textwidth]{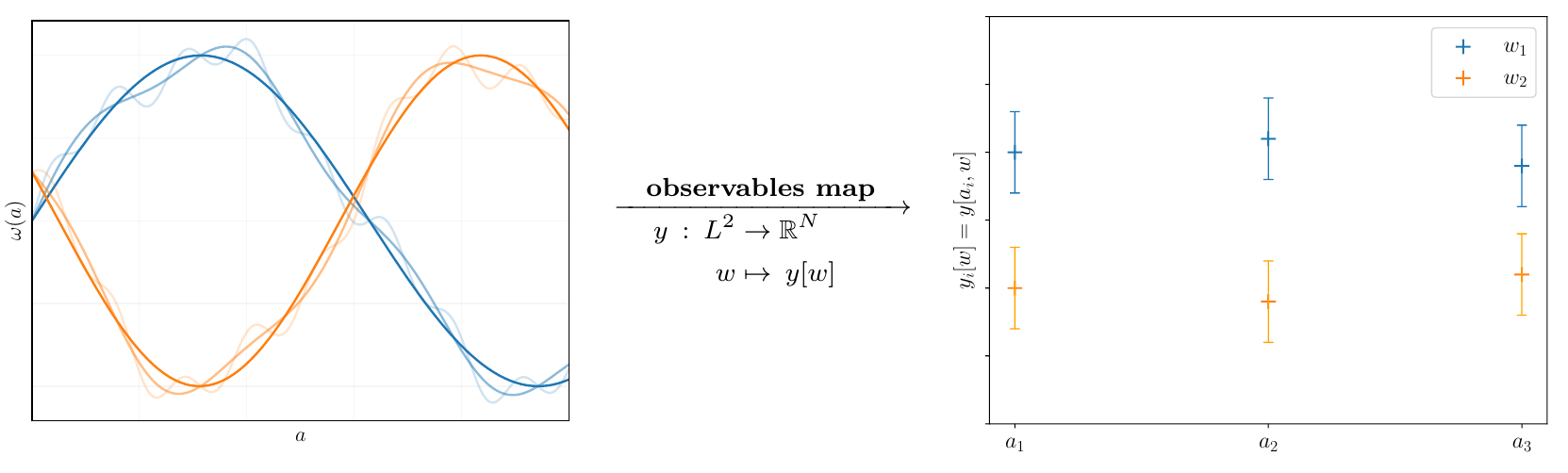}
    \caption{Illustration of the equivalence class construction. Left: two equivalence classes are given with three different model functions each. The functions are equivalent in the sense that they provide an identical prediction for $N$ observables within a certain allowed degree of difference is depicted by bars. Right: representatives $w_1, w_2$ of two different equivalence classes and their respective predicitions for three obervables.}
    \label{fig:space_cake_layers}
\end{figure}

This dimensionality issue can be resolved by the sensible restriction from the manifold of \emph{all} possible parameter functions $L^2(K)$ to the quotient space of those that the data allows us to distinguish, as illustrated for two equivalence classes by Fig.~\ref{fig:space_cake_layers}. A model $y: L^2(K) \rightarrow \mathbb{R}^{N}$ maps a parameter function $\w(a)$ to a set of $N$ predicted observations. One can define an equivalence relation $\sim$ on $L^2(K)$
\begin{equation}
    \w \sim \w' \quad \text{if and only if} \quad y[\w] = y[\w'],
\end{equation}
i.e. two parameter functions are equivalent if they yield the same predicted observation. Due to the inevitable presence of rounding errors and inexactness, this construction must be specified even further: two models will be considered equivalent if their predicitions for the $N$ observables \textit{all} agree within the a certain margin of difference. Consequently, $L^2(K)$ is partitioned into equivalence classes $[\w]$, which are in one-to-one correspondence with the possible observations, $\im y$. Now, by construction, we find \mbox{$\dim \nicefrac{L^2(K)}{\sim}  = \dim \im y \leq N$}, thus recovering the finite dimensional case. Therefore, one again reaches a non-degenerate Fisher metric on the quotient space. Note that a priori, this quotient space in general does not share all properties with $L^2(K)$: To check if the quotient space is again a manifold with a scalar product, one needs to check if the functional $y$ is open and continuous such that the induced quotient space is again separable. We will omit those technical details here, but it is worth noting that the distance modulus functional we will later employ, satisfies these requirements.

\subsubsection{Functional partition sums and the Gaussian path integral}
The functional partition sum follows from the case of discrete parameters $\theta^\mu$ in direct generalisation to continuous indexing $\w(a)$, leading to a path integral,
\begin{equation} \label{general def functional Z}
    Z[\beta,J_\alpha] = \int \mathrm{d}^n\theta\; \exp\bigg(-\frac{\beta}{2} \chi^2 (y,\theta) + J_\mu \theta^\mu\bigg) \quad \longrightarrow \quad 
    Z[\beta,J(a)] = \int \mathrm{D}\w \; \exp\bigg(-\frac{\beta}{2} \chi^2 [y,\w(a)] + \int \mathrm{d}a \: J(a) \w(a)\bigg).
\end{equation}
which itself is a functional, as it depends on the source functions $J(a)$ instead of a discrete set of $J_\alpha$, and the integration over discrete parameters becomes a functional (or path) integration
over the space of functions $\w(a)$. 

Unfortunately, path integrals are undefined in most instances. If we presume that the construction of
equivalence classes as described above is given, we can write $F_{a a'} = F[\w(a), \w(a')]$ and let $F^{a a'}$ denotes the inverse of the Fisher functional. Then, the Gaussian path integral presents a rare case that does
have an analytic solution, given by
\begin{equation}
     \frac{\int \mathrm{D}\w \: \exp\bigg(-\frac{\beta}{2}\int\mathrm{d}a \int \mathrm{d}a'\: F_{aa'} \:\w(a) \w(a')+\int \mathrm{d}a \: J(a) \w(a)\bigg)}{\int \mathrm{D}\w\:\exp\bigg(-\frac{\beta}{2}\int\mathrm{d}a\int \mathrm{d}a'\: F_{aa'} \:\w(a) \w(a')\bigg)}= \exp\bigg(\frac{1}{2 \beta} \int \mathrm{d}a \int \dd a' \:F^{aa'} J(a)  J(a')\bigg).
     \label{eq:1}
\end{equation}
The suitable normalisation is acquired automatically in the calculation of the
posterior's cumulants \citep[][]{roever2022partition}, due to the logarithm of the partition function, e.g. for
the second cumulant in the Gaussian case one finds through functional differentiation of $Z[\beta,J(a)]$
\begin{equation}
    \frac{\delta^2 \ln Z}{\delta J(a) \delta J(a')}\bigg|_{J=0} = 
    \frac{\delta}{\delta J(a')}\bigg[\frac{1}{Z} \frac{\delta Z}{\delta J(a)}\bigg]\bigg|_{J=0} = 
    -\frac{1}{Z^2}  \frac{\delta Z}{\delta J(a)} \frac{\delta Z}{\delta J(a')}\bigg|_{J=0} + \frac{1}{Z}  \frac{\delta^2 Z}{\delta J(a) \delta J(a')}\bigg|_{J=0}, 
\end{equation}
where the first term vanishes when the first functional derivative of $Z$ is evaluated at $J=0$. For the Gaussian case one finds
\begin{equation}
\frac{1}{\beta} \frac{\delta^2 \ln Z}{\delta J(a) \delta J(a')}\bigg|_{J=0} = 
\frac{1}{\beta} \frac{\delta^2}{\delta J(a) \delta J(a')} \exp \bigg(\frac{1}{2 \beta} \int \dd a \int \dd a' \:  F^{aa'} J(a) J(a')\bigg)\bigg|_{J=0} = 
\beta^{-2} \:F^{aa'} \xrightarrow{\beta \to 1} F^{aa'}
\label{eq:2},
\end{equation}
for the second cumulant of the functional posterior distribution. The first cumulant vanishes as the mean of the data was implicitly set to $y=0$ in eqn.~\eqref{eq:1} for simplicity. At this point it becomes evident that a degenerate Fisher metric corresponds to an infinite uncertainty in the parameters, thus inhibiting their inference. However, as discussed previously, there are ways to construct an equivalent of an inverse metric for non-singular manifolds, which may allow the use of the inverse Fisher-matrix in a sensible way.

\subsubsection{Estimates of functions and the associated variance for linear functional models}
\label{sec:linear_model_blue}
In general, the likelihood $\likeli(-\chi^2/2)$ of a measurement with a Gaussian error process takes the form
\begin{equation}
    \chi^2 = \bigg( y^i - y^i_{\text{model}}\bigg) \:C_{ij} \:\bigg( y^j - y^j_{\text{model}}\bigg),
    \label{chi_2}
\end{equation} 
with $C_{ij}$ as the inverse data covariance. A linear model $y^i_{\text{model}} = A\indices{^i_\mu}\theta^\mu$ depending on discrete parameters $\theta^\mu$, which becomes in the case of a functional dependence of the model $A^i(a)$. For notational simplicity, we write the data tuple as a linear form $y_i \equiv C_{ij} y^j$ as it would result by contracting the data vector with the inverse covariance matrix, such that it contains the measurement error. Then, the $\chi^2$-functional reads
\begin{equation}
    \chi^2 = 
    y_iy^i -
    2\int \mathrm{d}a \: y_i A^i(a) \:\w(a) + 
    \int\mathrm{d}a \int \mathrm{d}a' \: A_i(a)A^i(a')\: \w(a)\w(a'),
    \label{eq:chi2linear}
\end{equation}
leading to
\begin{equation}
\chi^2 = -2\int\mathrm{d}a\:Q_a\w(a) + \int\mathrm{d}a\:\int\mathrm{d}a'\: F_{aa'}\: \w(a)\w(a')
\quad\text{with}\quad
Q_{a} = y_i A^i(a)
\quad\text{and}\quad
F_{aa'} = A_i(a)A^i(a').
\end{equation}
In analogy to the discrete parameter case, the lower Cramér-Rao bound is reached with a Gaussian functional likelihood \citep[][for a derivation proceeding over a partition function]{roever2022partition}, the second cumulant as given in \eqref{eq:2},
\begin{equation}
    \frac{\int \mathrm{D}\w \: \exp\bigg(-\frac{\beta}{2}\int\mathrm{d}a \int \mathrm{d}a'\: F_{aa'} \:\w(a) \w(a')+\int \mathrm{d}a \: J(a) \w(a)\bigg)\: w(a) w(a')}{\int \mathrm{D}\w\:\exp\bigg(-\frac{\beta}{2}\int\mathrm{d}a\int \mathrm{d}a'\: F_{aa'} \:\w(a) \w(a')\bigg)}\Bigg|_{J = 0} = \langle \w(a) \w(a') \rangle = F^{aa'}.
\end{equation} 
The Gaussian likelihood thus minimizes the parameter variance, providing the estimate with highest possible precision for $\w$. In the discrete case, the unbiasedness of a Gaussian likelihood articulates itself as $\bar{\theta}^\mu = \hat{\theta}^\mu $ with $\bar{\theta}^\mu$ being the best fit parameter and $\hat{\theta}^\mu$ the true value. To compute the best-fit function $\bar{\w}(a)$, a functional minimisation
of $\chi^2$ is performed with respect to $\w(a)$,
\begin{align}
    \frac{\delta \chi^2}{\delta \w(a)} = -2 Q_a + 2 \int \mathrm{d}a' \:F_{aa'} \:\w(a') \stackrel{!}{=}0,
\end{align}
using the symmetry of the Fisher-functional. Solving the resulting condition for the best estimate $\bar{\w}(a)$ yields
\begin{equation}
     Q_a = \int\mathrm{d}a\: F_{aa}\:\bar{\w}(a) 
     \quad \rightarrow \quad 
     y^i = \int \mathrm{d}a \: A^i(a) \:\bar{\w}(a). 
\end{equation}
As the data stems from a true model $\hat{\w}(a)$ and the data is generated through $y^i = \int \mathrm{d}a \: A^i(a) \hat{\w}(a)$, this leads to 
\begin{equation}
    \int \mathrm{d}a \: A^i(a) \:\hat{\w}(a) = \int \mathrm{d}a \: A^i(a)\: \bar{\w}(a). \label{eq:4}
\end{equation}  
Hence, the unbiasedness of the Gaussian likelihood in the functional case means that the integrals of the functions over their domain must coincide. As this equality must hold for all domains, one can indeed conclude equality of the integrands $\hat{\w}(a) = \bar{\w}(a)$ from \eqref{eq:4}. Furthermore, \eqref{eq:4} underlines the fact that in the functional case, what is of actual relevance for generating the data is the \emph{accumulated effect} of the function $\w$ evaluated over its entire domain.

\subsubsection{Functional information entropy}
The information entropy can be derived from the partition sum as follows \citep[][]{roever2022partition},
\begin{equation}
    S = -\beta^2 \frac{\partial}{\partial \beta} \bigg(\frac{1}{\beta} \ln Z\bigg)\bigg|_{\beta = 1} = \ln Z -  \frac{\beta}{Z} \frac{\partial Z}{\partial \beta}\bigg|_{\beta = 1}.
\end{equation}
In general, Gaussian path integrals are only defined with a suitable
normalisation, as explained previously. The first term in the expression for
the information entropy, $\ln Z$, does not naturally provide such a
normalisation. To circumvent this, we simply set $Z|_{\beta = 1} = 1$, falling back onto the Bayes-evidence, and find
\begin{equation}
    S = 
    -\left (\frac{\int  \mathrm{D}\w \: \exp\bigg(-\frac{\beta}{2} \int\mathrm{d}a\int\mathrm{d}a'\: \:F_{aa'} \:\w(a) \w(a')\bigg) \;\bigg(-\frac{1}{2} \int \mathrm{d}a \int \mathrm{d}a'\:F_{aa'}\: \w(a) \w(a')\bigg)}{\int  \mathrm{D}\w \: \exp\bigg(-\frac{\beta}{2} \int\mathrm{d}a\int\mathrm{d}a'\:F_{aa'} \:\w(a) \w(a')\bigg)}\right)_{\beta = 1} = 
    - \frac{1}{2}\int \mathrm{d}a \int \mathrm{d}a'\:F_{aa'}\:F^{aa'},
\end{equation}
which diverges for a functional Fisher metric. This is expected since a function taking on the role of a random variable admits an infinite information content. In the degenerate functional case, the information
entropy is not defined (due to the non-existence of the inverse Fisher metric), but this does not necessarily mean that the functional approach is not viable. When the space of functions is limited to equivalence classes of functions agreeing on a finite number of observables (cf. \ref{subsubsection:fisher-functional}), the information entropy remains finite.

\section{Dark Energy Equation of State from Type Ia supernovae measurements}\label{sect_supernova}
As a topical example of cosmology we try out functional inference with the dark energy equation of state as a function of scale factor $a$, constrained by supernova data. It will be assumed that this function does not deviate much from the well-justified $\Lambda$CDM-prediction $\w\equiv -1$. Thus, effectively, a dark energy equation of state of the form $\w(a) = -1 + \delta \w(a)$ is investigated with variations $\delta\w$ around the fiducial $\Lambda$-case.

Canonical sampling with a Monte Carlo Markov-chain will be used to constrain this function for the case of a linear model, and later for a model that is quadratic in the parameters. The data in consideration are measurements of distance moduli from type Ia supernovae (\citet{Suzuki_2012}, \citet{Kowalski_2008}, \citet{amanullah_spectra_2010}). If one allows the dark energy equation of state $\w (a)$ to vary with cosmic time, the distance modulus at scale factor $a_{i}$ is given by \citep[][]{Hubble_de_eos_takada},
\begin{equation}
    y[a_i,\w] = 
    5\log \bigg[\frac{1}{a_i} \int_{a_i}^1 \mathrm{d}x' \: \frac{1}{{x'}^2\: H_0 \sqrt{\Omega_m x'^{ -3}+\Omega_\mathrm{DE}\exp\Big(-3 \int_1^{x'} \mathrm{d}x \frac{(1+\w(x))}{x}\Big) }}\bigg]+10.
\end{equation}
with the constraint of spatial flatness requiring $\Omega_m + \Omega_\mathrm{DE} = 1$. As an example, we set up a Gaussian likelihood for a supernova data sample \citep{goobar_supernova_2011} with the simplifying assumption of uncorrelated data. For computationally intensive tasks, we use physics-informed neural networks to generate fast and accurate predictions of $y[a,\w]$ \citep{roever2022partition, 2023arXiv230507061R}.

\subsection{Results for the linear case}\label{Results for the linear case}
\begin{figure}
	\centering
    \includegraphics[width=0.7\textwidth]{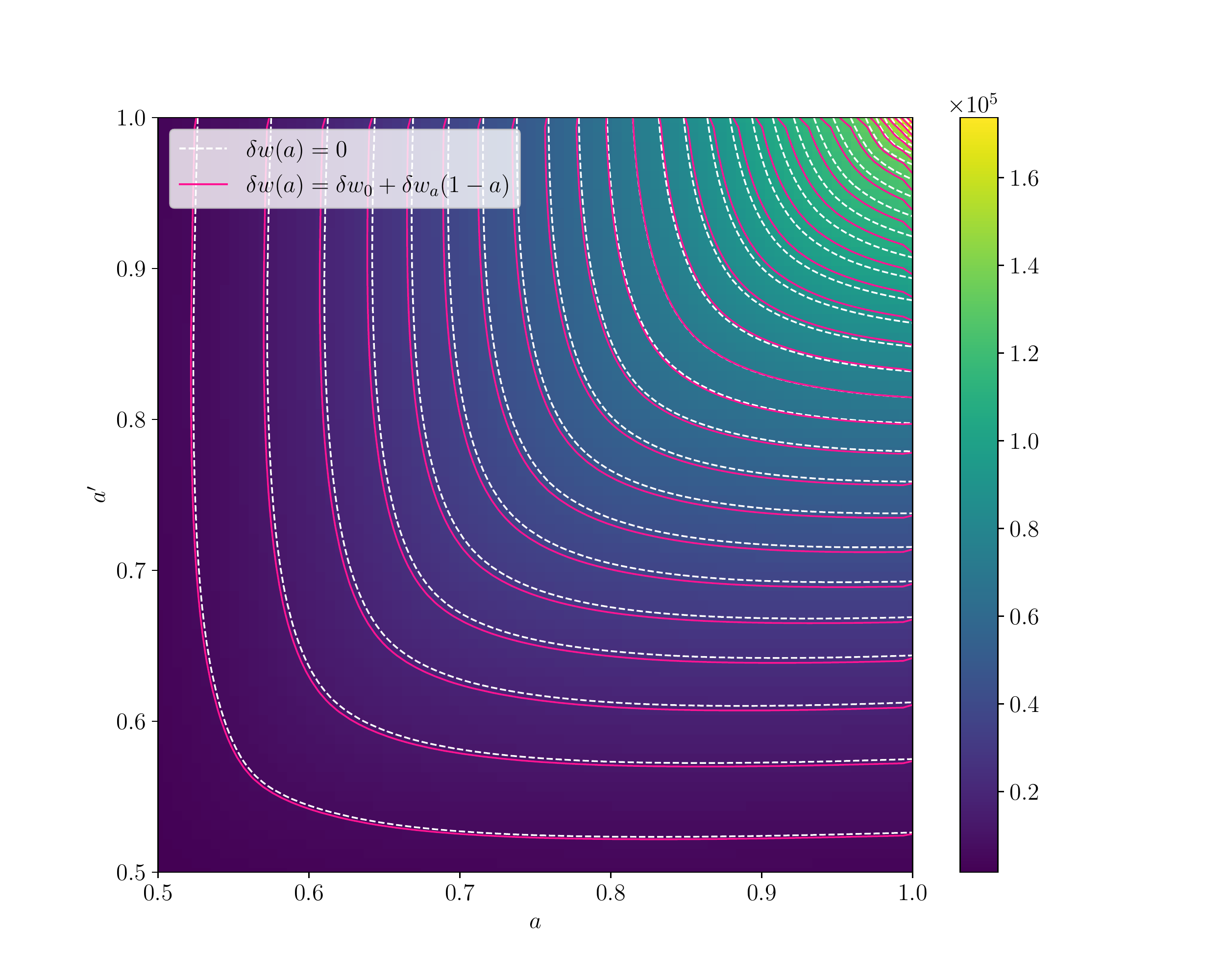}
    \caption{Fisher-functional $F[\w(a) \w(a')]$ at the fiducial equation of state $\w(a)=-1$ (recovering the linear model with $\delta \w(a)=0$) and at an exemplary Chevallier-Polarski-Linder model ($\w(a) = -1 + \delta \w_0 + \delta \w_a \: (1-a)$ with $\w_0=0.8$ and $\delta \w_a=0.3$) as functions of the scale factors.}
    \label{fig_fisher_functional}
\end{figure}

To begin with, a linear model with a Gaussian likelihood as in Sect.~\ref{sec:linear_model_blue} is considered, hence the distance modulus ought to be linearised in the function $\delta \w(a)$. The distance modulus is linearised in the function $\delta \w(a)$ using a Volterra series as the functional pendant of a Taylor-expansion, assuming the functional perturbations $\delta \w$ to be small at all cosmic times. 
\begin{equation}
    y[a_i,\w] \cong 
    y[a_i, \delta \w = 0 ] + 
    \int\mathrm{d}a\: \frac{\delta y[a_i, \w]}{\delta \w(a)}\bigg|_{\w(a) \equiv -1} \delta\w(a) + \mathcal{O}[(\delta \w)^2],
\end{equation}
with, $y[a_i, \delta \w = 0] = y[a_i, \w = -1]$ is the distance modulus as given by the $\Lambda$CDM model. The first order term provides a functional correction to $\Lambda$CDM. One finds
\begin{equation} \label{eq:scalefactor_expansion}
y[a_i,\w] \cong 
y[a_i, \w=-1] - K(a_i)  \int_{a_i}^1 \mathrm{d}x'\; \frac{1}{{x'}^2 (\Omega_m {x'}^{-3} \Omega_\mathrm{DE})^{3/2}} \int_{x'}^1 \mathrm{d}a\:\frac{\delta \w(a)}{a}
\end{equation}
where we refer to Appendix~\ref{appendix_functional_derivatives} for details concerning functional derivatives, with
\begin{equation}
K(a_i)\equiv 
\frac{15}{2 \ln(10)}\:\Omega_\mathrm{DE} \:\bigg(\int_{a_i}^1 \mathrm{d}x' \frac{1}{{x'}^2 \sqrt{\Omega_m {x'}^{-3}+\Omega_\mathrm{DE}}}\bigg)^{-1}.
\end{equation}
The Fisher-functional $F_{aa'}$ for the dark energy equation of state function $\w(a)$ constrained by supernova data is shown in Fig.~\ref{fig_fisher_functional}, specifically for a linear and a quadratic model. As opposed to the linear case, the Fisher-functional for the quadratic model possesses a dependence on $\delta \w(a)$ (for a discussion of the quadratic model, we refer to section \ref{Results for the quadratic case}). Thus, an exemplary function $\delta \w(a)$ must be chosen. The great similarity of the two figures suggests that the effect of higher-order terms on the Fisher-functional is rather small in this example. Large values of $F_{aa'}$, corresponding to high constraining power, are obtained at values of $a$ close to unity, and we truncate the matrix at values of $a\simeq 1/2$, as the supernova data becomes sparse at redshifts above one.

For the reasons outlined in the previous chapters, one needs to return to a finite number of parameters $N$ in order to actually infer parameters of a physical model. In the following, it is shown that a Fourier series expansion of the functions in question indeed allows to recover the finite case. Moreover, introducing a cutoff-frequency imposes a smoothness scale on the function to be inferred. This cutoff relates back to the equivalence class construction, as the marginalization undergone by the function in the distance modulus might cause the effect of higher-order modes to be negligible for the prediction of observables. As two functions are considered to be equivalent when their predicitions coincide, expansions of functions with any number of permitted modes may fall into the same equivalence class. Going forward, Occam's razor seems advisable when selecting a function in such a case. Periodicity of $\w(a)$ can be assumed by simply considering the relevant cosmic time interval $[0,1]$ to cover \textit{half} a period of the function, thus not imposing any additional constraints on the function. The same is assumed for $J(a)$. Cosines are chosen as a functional basis, 
\begin{equation}
    \w(a) = \frac{\w_0}{2} + \sum_{n=1}^\infty \w_n \cos(\pi n a); \quad J(a) = \frac{J_0}{2}+\sum_{n=1}^\infty J_n \cos(\pi n a).
\end{equation}
The functional canonical partition function \eqref{general def functional Z} thus becomes, 
\begin{align}
       Z [\beta, J(a)] &= \int \mathrm{D}\w \: \exp\Big(-\frac{\beta}{2}\int \mathrm{d}a \int \mathrm{d}a'\: F_{aa'} \:\Big(\frac{\w_0}{2} + \sum_{n=1}^\infty \w_n \cos(\pi n a) \Big)\Big(\frac{\w_0}{2} + \sum_{n'=1}^\infty \w_{n'} \cos(\pi n' a')\Big)\Big) \nonumber \\
       &\hspace{1.15cm}\times \exp\Big(\int \mathrm{d}a \: \Big(\frac{J_0}{2}+\sum_{n=1}^\infty J_n \cos(\pi n a)\Big) \Big(\frac{\w_0}{2} + \sum_{n'=1}^\infty \w_{n'} \cos(\pi n' a)\Big)\Big).
       \label{Functional Z}
\end{align}
For the case of a linear model with Gaussian likelihood, the expression for the Fisher-functional on the interval $[0,1]$ was identified in \eqref{eq:chi2linear} in terms of the functions $A^i(a)$. These functions can be expanded in Fourier series as well with coefficients $\Tilde{A}\indices{^i_n}, \Tilde{A}\indices{^i_{n'}}$. This allows us to read off the Fisher information in discrete frequency space $(F_{n n'})_{n, n' \in \mathbb{N}}$ (for a detailed derivation see Appendix~\ref{Fourier Coefficients}). The functional partition sum in the Fourier-representation then reads
\begin{equation}
    Z [\beta, J_n] = \int \mathrm{d}^n\w \: \exp\bigg(-\frac{\beta}{2} \bigg(F_{00}\frac{\w_0^2}{2} + \sum_{n = 1}^\infty F_{0n} \frac{\w_0}{2}\w_n + \sum_{n = 1}^\infty \sum_{n' = 1}^\infty F_{nn'}\w_n \w_{n'} + \frac{1}{2} J_0 \frac{\w_0}{2} + \frac{1}{2} \sum_{n = 1}^\infty J_n \w_n \bigg)\bigg).
\end{equation}
As long as an infinite number of Fourier modes is permitted, the Fisher matrix is still infinite dimensional. When a cut-off $N$ is introduced, this result coincides with the familiar case of a finite number of parameters, as expected. Applying this idea now to the dark energy equation of state by expanding the (non-periodic) perturbation in Fourier modes,

\begin{equation}
    \delta \w(a) = \frac{\delta \w_0}{2} +\sum_{n=1}^\infty \: \delta\w_n \cos(\pi n a),
    \label{perturbation fourier}
\end{equation}
and inserting it in \eqref{eq:scalefactor_expansion}, yields
\begin{align}
     y[a_i,\w] & 
     \cong y[a_i, \w = -1 ] + \delta \w_0 \frac{K(a_i)}{2} \bigg( \int_{a_i}^1 \mathrm{d}x' \: \frac{\ln(x')}{{x'}^2 (\Omega_m {x'}^{-3}+\Omega_\mathrm{DE})^{3/2}} \bigg)- \sum_{n=1}^N \: \delta\w_n  K(a_i)   \bigg(\int_{a_i}^1 \mathrm{d}x'\; \frac{[\mathrm{Ci}(\pi n) -\mathrm{Ci}(\pi n x')]}{{x'}^2 (\Omega_m {x'}^{-3}+\Omega_\mathrm{DE})^{3/2}} \bigg)\nonumber \\
     &\equiv y[a_i, \w = -1 ] + \Delta y[a_i, \delta \w].
     \label{eq: AGauss}
\end{align}
Now let $y^i = y(a_i)$ be the measured SNIa distance modulus at scale factor $a_i$, which are assumed to be statistically independent, for simplicity. Written in terms of the measured deviations from $\Lambda$CDM,
$\Delta y^i \equiv y^i - y[a_i, \w = -1]$, one finds with \eqref{chi_2}, again writing $y_i = C_{ij} y^j$
\begin{equation}
    \chi^2 = 
    \Big(\Delta y_i - \sum_{n=0}^N A\indices{_{i,n}} \:\delta\w_n \Big)\Big(\Delta y^i - \sum_{n'=0}^N A\indices{^i_{n'}} \:\delta\w_{n'} \Big) = 
    \Delta y_i\Delta y^i- 
    2 \Delta y_i \sum_{n=0}^N A\indices{^i_n}\: \delta \w_n + 
    \sum_{n=0}^N \sum_{n'=0}^N A\indices{_{i,n}} A\indices{^i_{n'}} \:\delta \w_n \delta \w_{n'} \equiv 
    2\:\sum_{n=0}^N  Q_n\:\delta\w_n + \sum_{n=0}^N \sum_{n'=0}^N F_{n n'} \; \delta \w_n \delta \w_{n'},
\end{equation}
up to an irrelevant additive factor, and with $A\indices{^i_n}$ being the Jacobian needed to move from data space to parameter space for a linear model. By comparison with \eqref{eq: AGauss}, 
\begin{align}
    A\indices{^i_0} &= +K(a_i)   \bigg( \int_{a_i}^1 \mathrm{d}x' \: \frac{\ln(x')}{2{x'}^2 (\Omega_m {x'}^{-3}+\Omega_\mathrm{DE})^{3/2}} \bigg), \\
    A\indices{^i_m} &= -K(a_i)   \bigg(\int_{a_i}^1 \mathrm{d}x'\; \frac{[\mathrm{Ci}(\pi m) -\mathrm{Ci}(\pi m x')]}{{x'}^2 (\Omega_m {x'}^{-3}+\Omega_\mathrm{DE})^{3/2}} \bigg), \quad (\text{for} \; m\geq 1). 
\end{align}
Thus, the Fisher matrix components are given accordingly, for 580 data points from SN1a measurements assuming statistical independence,
\begin{equation}
    F_{nn'} = \sum_i \frac{A\indices{^i_n} A\indices{^i_{n'}}}{\sigma_i^2}.
\end{equation}

\begin{figure}
    \includegraphics[width = 10cm]{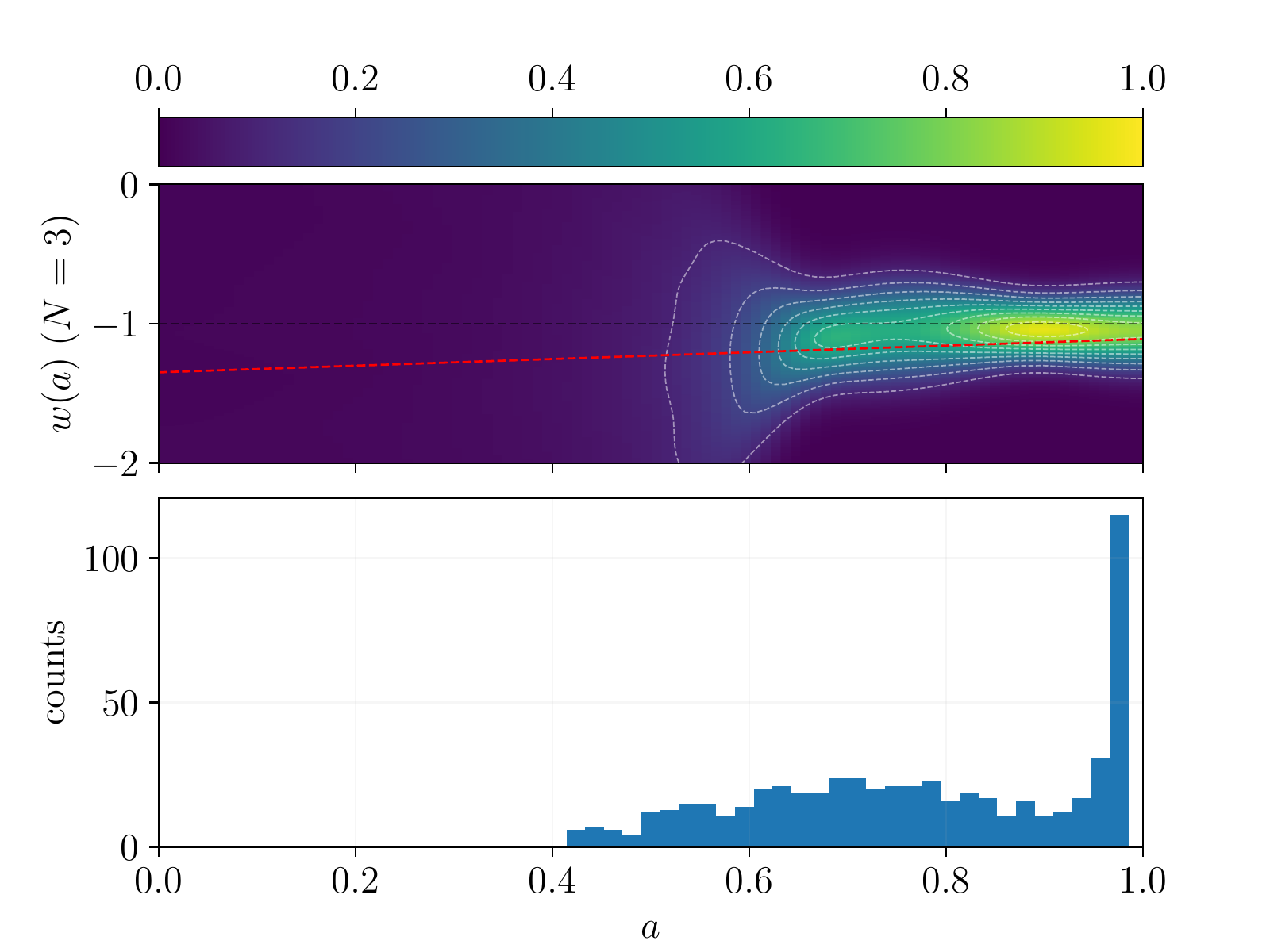}
    \centering
    \caption{The posterior distribution of the dark energy equation of state $\w(a)$ with 3 modes in the linear model is compared to the distribution of SNIa data points in scale factor $a$. Here and in the following, the dotted red line displays the Chevallier-Polarski-Linder (CPL) model with latest constraints for comparison. \citep{yang2018observational}}
    \label{3 modes linear}
\end{figure}

Subsequently, the Fourier coefficients in question can be inferred via MCMC sampling. The resulting posterior distribution of the dark energy equation of state is displayed in Fig.~\ref{3 modes linear} for a choice of three modes, in comparison with the number of SNIa data points available at different times in the cosmic time. Note that this the posterior distribution of a function, higher density thus indicates the most probable \textit{functional value $\w(a)$} at this particular scale factor $a$. Fig.~\ref{3 modes linear} nicely shows how the constraining power of the data decreases as the data becomes more sparse at earlier times. The average curve of the dark energy equation of state is given in pink.

\begin{figure}
    \includegraphics[width = 12cm]{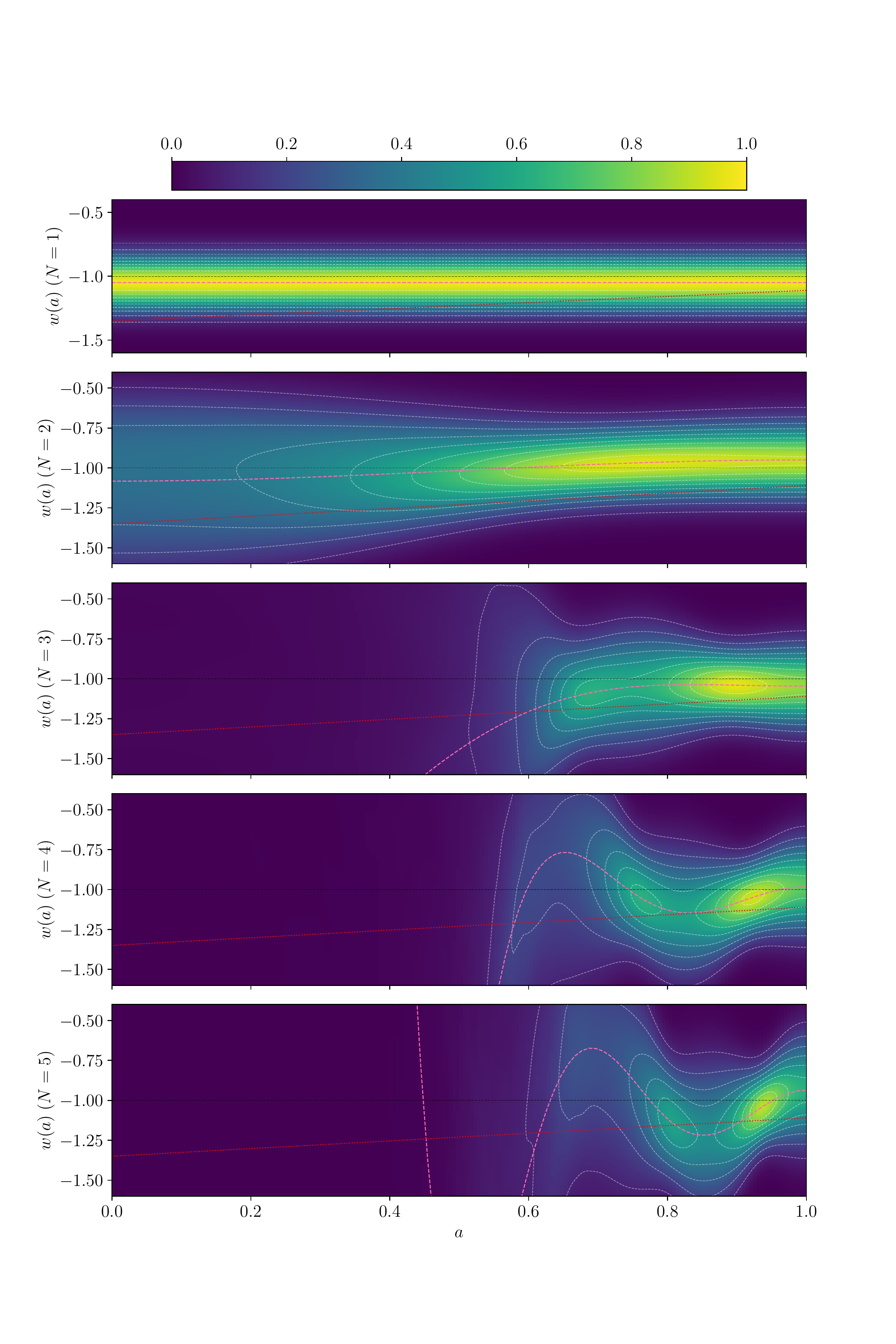}
    \centering\vspace{-1cm}
    \caption{The posterior distribution of the dark energy equation of state $\w(a)$ as a function of the scale factor for different choices of the number of modes $N$ in the linear model. The average curve is given in pink. Again, the dotted red line displays the CPL model for comparison.}
    \label{N modes linear}
\end{figure}

Fig.~\ref{N modes linear} illustrates the effect that an increasing complexity of the parameter space has on the predicted behaviour of the function. As expected, the posterior distribution of the dark energy equation of state starts to show more variability with cosmic time and sets in in the more recent past, as more Fourier modes are permitted. With increasingly many coefficients, the time interval for which the posterior function is well-constrained decreases. In a way, the function reflects that the data is not sufficiently constraining more complex models, and that the loss of predictive power sets in at lower redshifts. There seems to be a strong effect of the model choice on extrapolation towards high redshifts, as almost all models suggest phantom-like behaviour $w<-1$ at early times \citep{majerotto_supernovae_2004}.

\begin{figure}
	\centering
	\includegraphics[width =0.45 \textwidth]{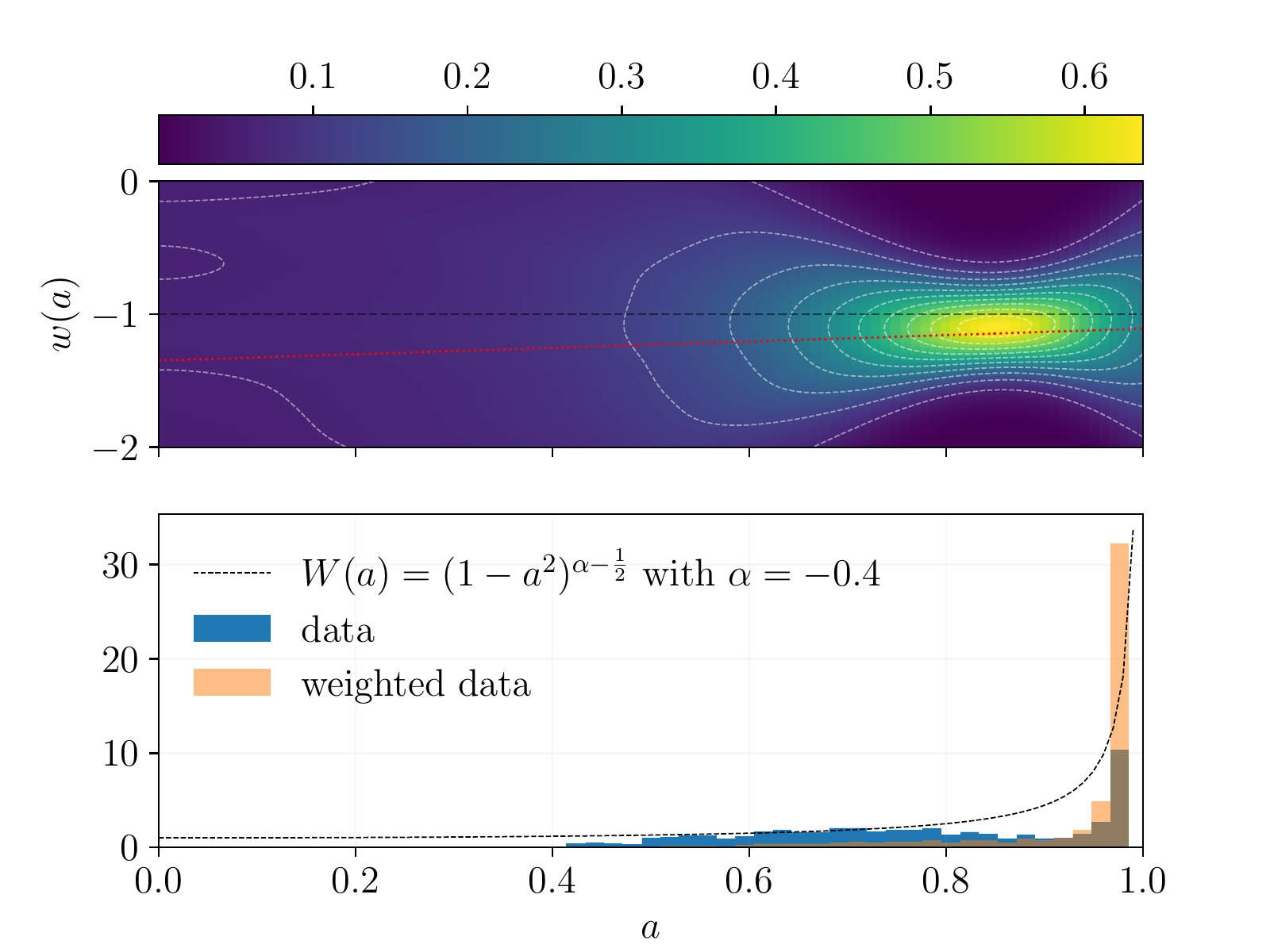}
	\includegraphics[width =0.45 \textwidth]{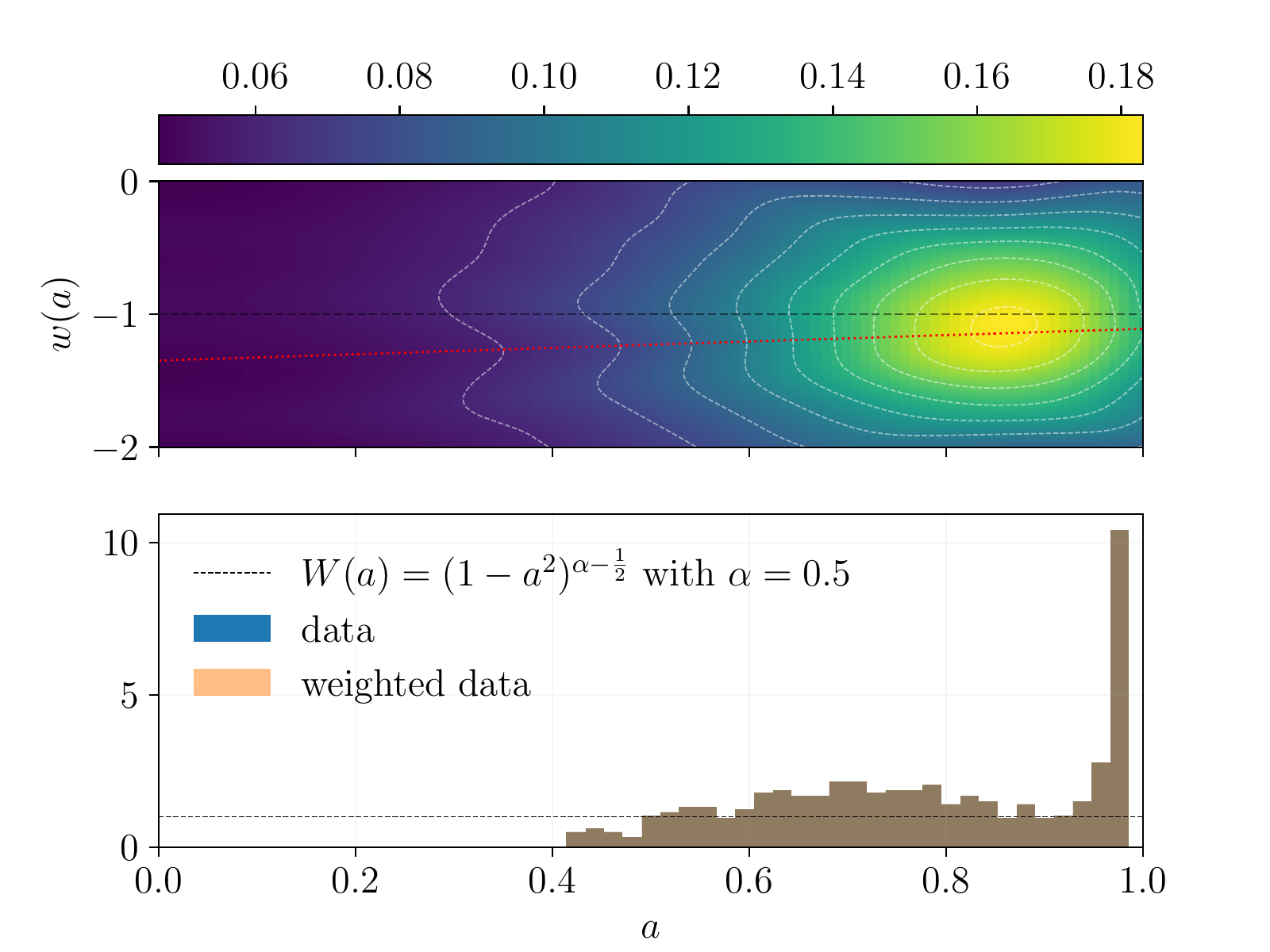}\\
	\includegraphics[width =0.45 \textwidth]{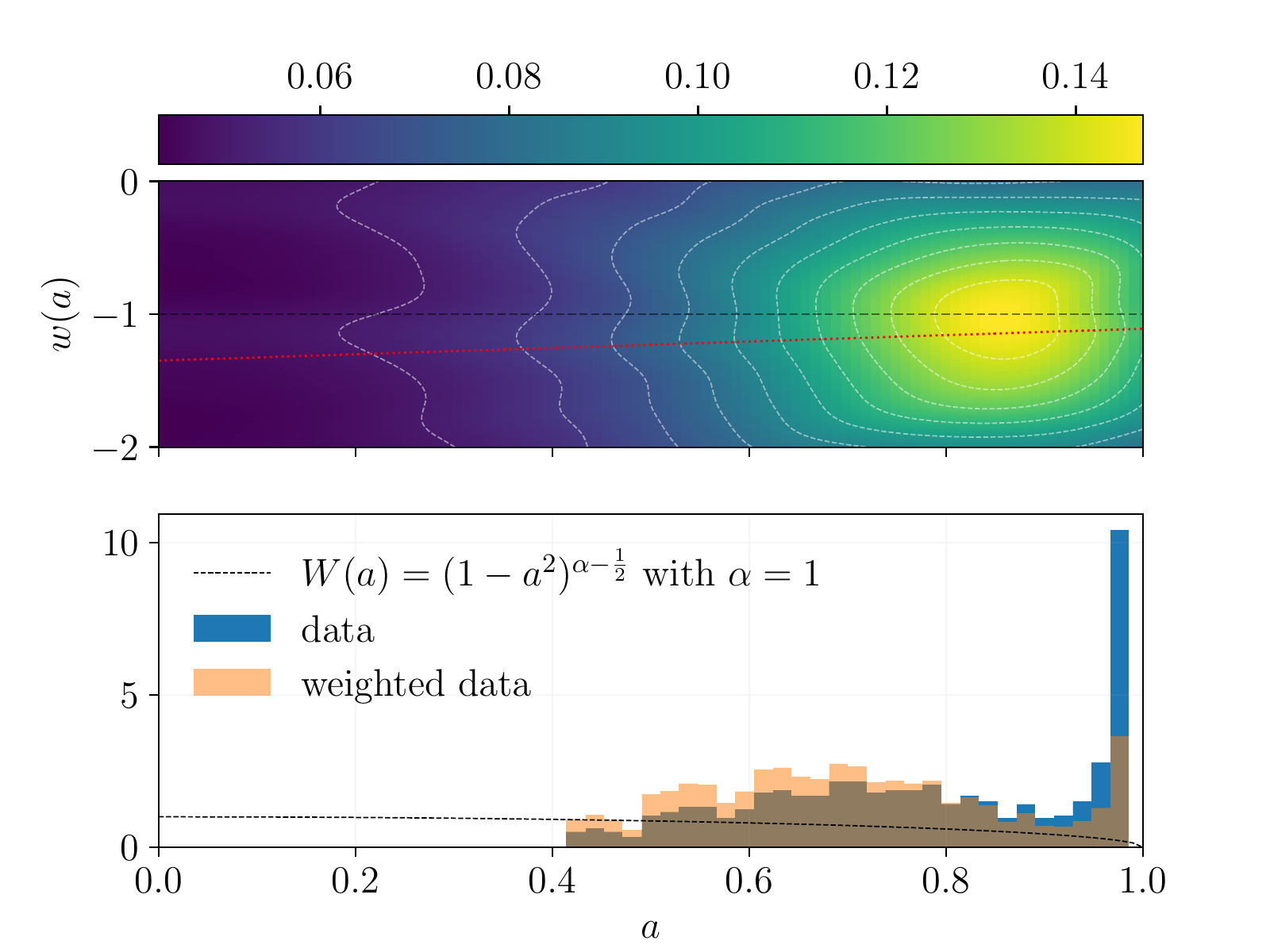}
	\includegraphics[width =0.45 \textwidth]{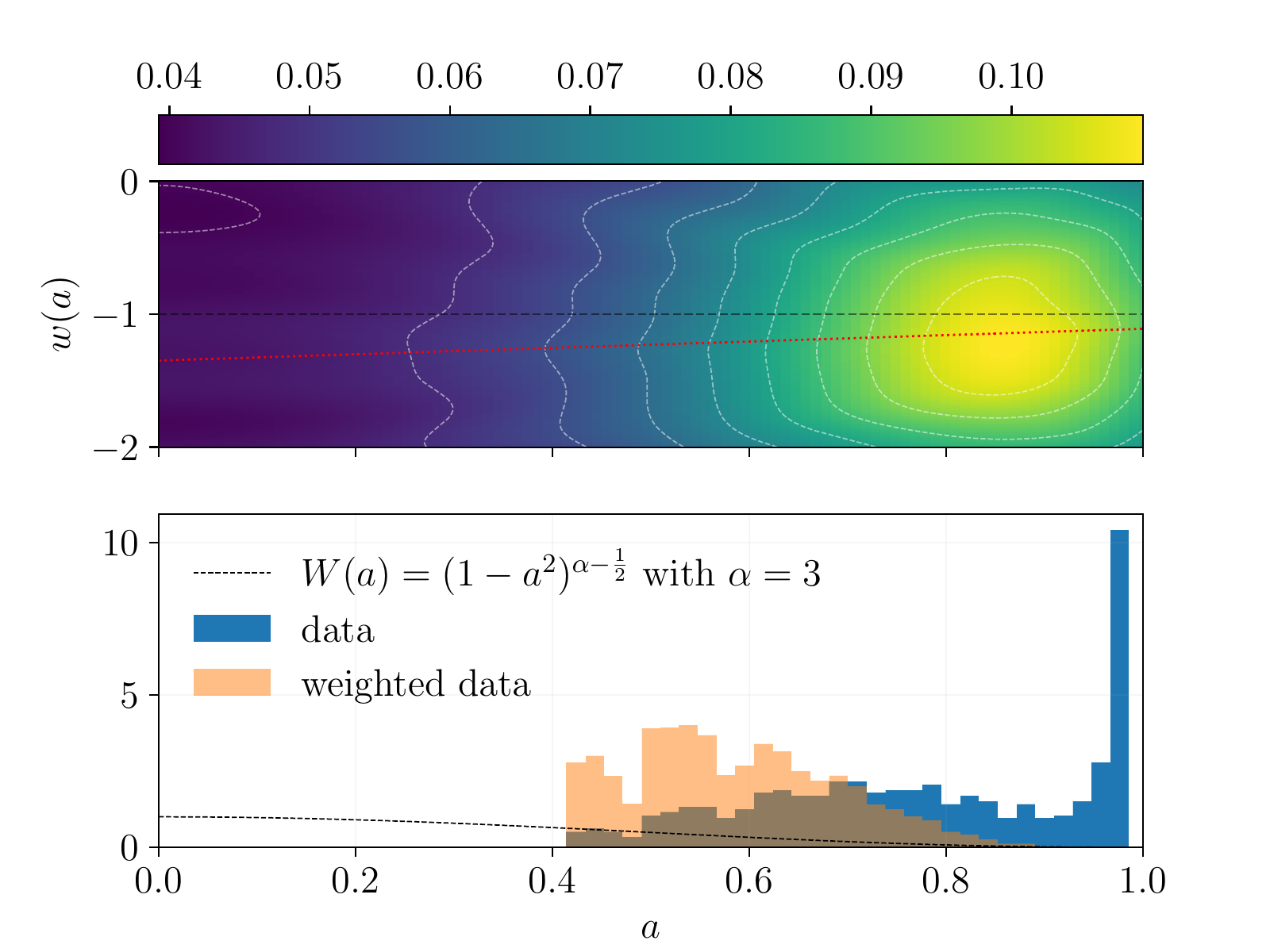}
    \caption{The posterior distribution of the dark energy equation of state $\w(a)$ is displayed in comparison to the distribution of SNIa data points and weighting as a function of the scale factor $a$. $N$ coefficients of the expansion in Gegenbauer polynomials were sampled for a linear model for three modes, for $\alpha = -0.4, 0.5, 1 ,3$. Again, the dotted red line displays the CPL model.}
    \label{gegenbauer}
\end{figure}

As an alternative to a Fourier series expansion, one can choose to expand in polynomials $P_n$ orthogonal on a closed interval with respect to a weighting function $W(a)$, i.e. $\smallint W(a) P_n(a) P_m(a) \propto \delta_{mn}$. This weighting function enables additional modifications. We choose the class of Gegenbauer polynomials, which have a weighting function $W(a) = (1 - a^2)^{\alpha - \frac{1}{2}}$ where $\alpha \in [ - 1,1]$ \citep{abramowitz1988handbook}. Here, the free parameter $\alpha \in \mathbb R$ permits putting emphasis on data at a certain scale factor, thereby increasing or decreasing its constraining power at certain cosmological epochs. Gegenbauer polynomials appear to be a particularly well-suited choice, as they form an infinite basis for the Hilbert space under consideration, just like the trigonometric functions do. Evidently, it is desirable to expand the $L^2$-function $\w(a)$ into a polynomial basis that spans $L^2([0,1])$. This weighting is visualised in Fig.~\ref{gegenbauer}, where the posterior distribution of $\w(a)$ is compared not only to the number of data points available at the respective scale factor, but also to the weighting function of the Gegenbauer polynomials for different values of $\alpha$. The set of Legendre-polynomials is recovered for the specific choice $\alpha = 0$.

\subsection{Results for the quadratic case}\label{Results for the quadratic case}
In order to arrive at a case with proper non-Gaussian distributions, we turn towards a model that is \emph{quadratic} in its parameters, i.e.
\begin{equation}
    y^i_\text{model} = 
    A\indices{^i _\mu} \theta^\mu + \frac{1}{2} B\indices{^i_{\mu \nu}} \theta^\mu \theta^\nu.
    \label{AB}
 \end{equation}
Let us again assume that the data are uncorrelated, that is, that their covariance takes a diagonal form, and the choice of units sets the diagonal entries to one. For a Gaussian error process, this may always be achieved by an affine transform. With this, the Fisher information reads
\begin{equation}
    F_{\mu\nu} = 
    A\indices{^i_\mu} A\indices{_{i,\nu}} + 2 A\indices{^i_{\nu}} B\indices{_{i,\mu\lambda}}\theta^\lambda
    + B\indices{^i_{\lambda\mu}} B\indices{_{i,\nu \kappa}} \theta^\lambda \theta^\kappa.
\end{equation}
Repeating the expansion as in \eqref{AB} in terms of powers of $A\indices{^i_\mu}$ and $B\indices{^i_{\mu\nu}}$) leads to 
\begin{equation} \label{eq:ABgauss}
\log\mathcal{L} \sim
\bigg(y^i - A\indices{^i_\mu} \theta^\mu - \frac{1}{2} B\indices{^i_{\mu\nu}}\bigg) C_{ij} \bigg(y^j - A\indices{^j_\mu} \theta^\mu - \frac{1}{2} B\indices{^j_{\mu\nu}}\bigg)
    \stackrel{y^i = 0}{\longrightarrow}\; 
    -\bigg(A\indices{^i_\mu} A\indices{_{i\nu}} \: \theta^\mu \theta^\nu + A\indices{^i_\lambda} B\indices{_{i,\mu \nu}} \theta^\lambda \theta^\mu \theta^\nu + \frac{1}{4}B\indices{^i_{\mu \nu}}B\indices{_{i,\lambda \delta}}\:\theta^\mu \theta^\nu \theta^\lambda \theta^\delta\bigg) 
\end{equation}
effectively providing a link to the DALI formalism \citep{sellentin_breaking_2014}, where the logarithmic likelihood is likewise expanded into powers of $\theta$ to avoid positivity issues
\begin{equation}
    \log \mathcal{L} \propto  
    -\frac{1}{2}F_{\alpha\beta}\:\theta^\alpha \theta^\beta 
    -\frac{1}{3!}S_{\alpha\beta\gamma} \:\theta^\alpha \theta^\beta \theta^\gamma 
    -\frac{1}{4!}Q_{\alpha\beta\gamma\delta} \:\theta^\alpha \theta^\beta \theta^\gamma \theta^\delta.
\end{equation} 
Here, the expansion was implicitly performed around the best fit point and $\theta^\alpha$ measured relative to that point. $S_{\alpha\beta\gamma}$ is refered to as the flexion, $Q_{\alpha\beta\gamma\delta}$ as the quarxion introduced by \citet{sellentin_breaking_2014}. Direct comparison for the two expansion schemes up to fourth order in the parameters results in $S_{\alpha\beta\gamma} = 3A\indices{^i_\alpha} B\indices{_{i,\beta\gamma}}$ and $Q_{\alpha\beta\gamma\delta} = 4B\indices{^i_{\alpha\beta}} B\indices{_{i,\gamma\delta}}$.

\subsection{Dark energy equation of state at quadratic order}
\begin{figure}
	\centering
	\includegraphics[width =0.45 \textwidth]{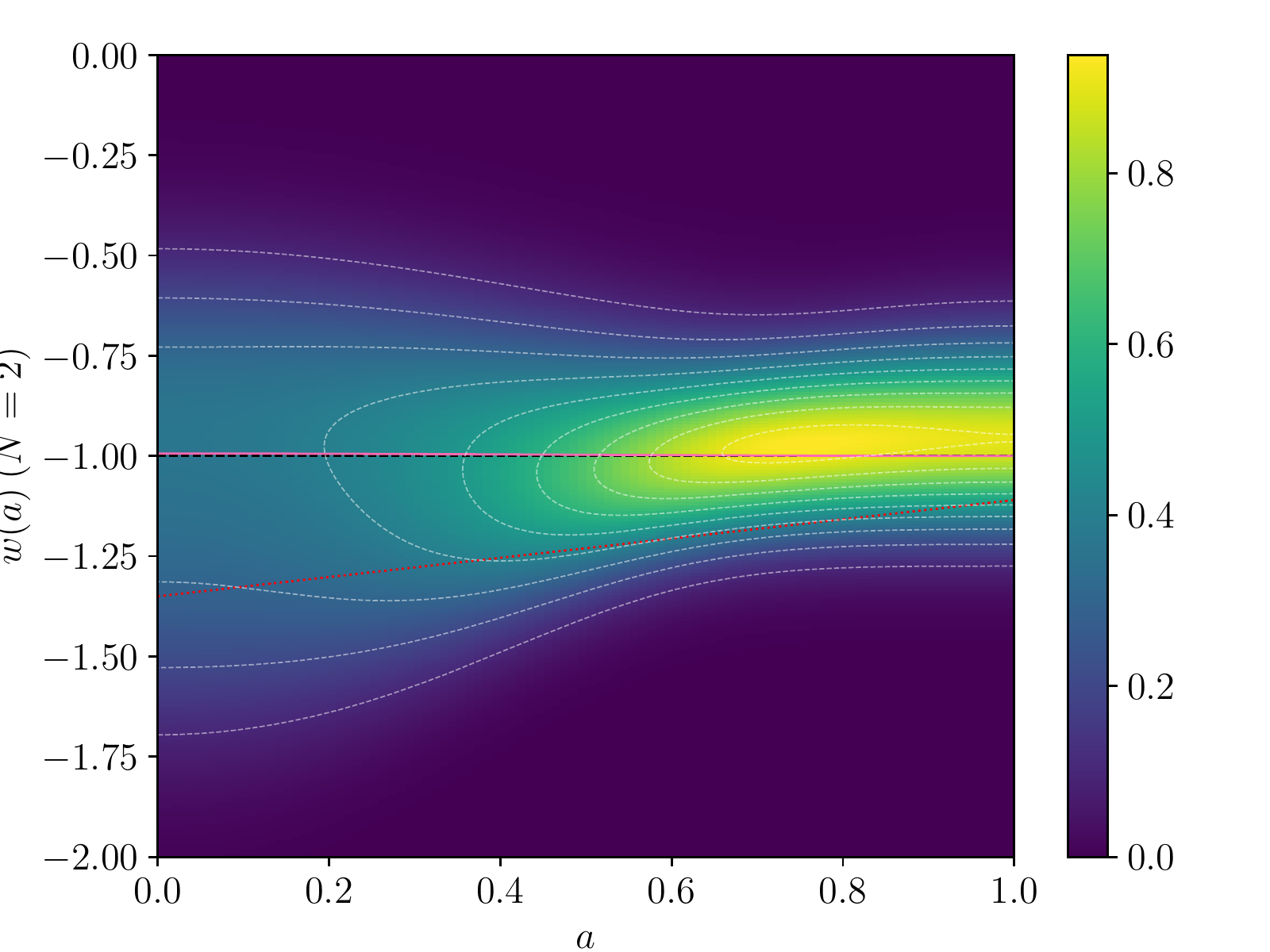}
    \includegraphics[width =0.45 \textwidth]{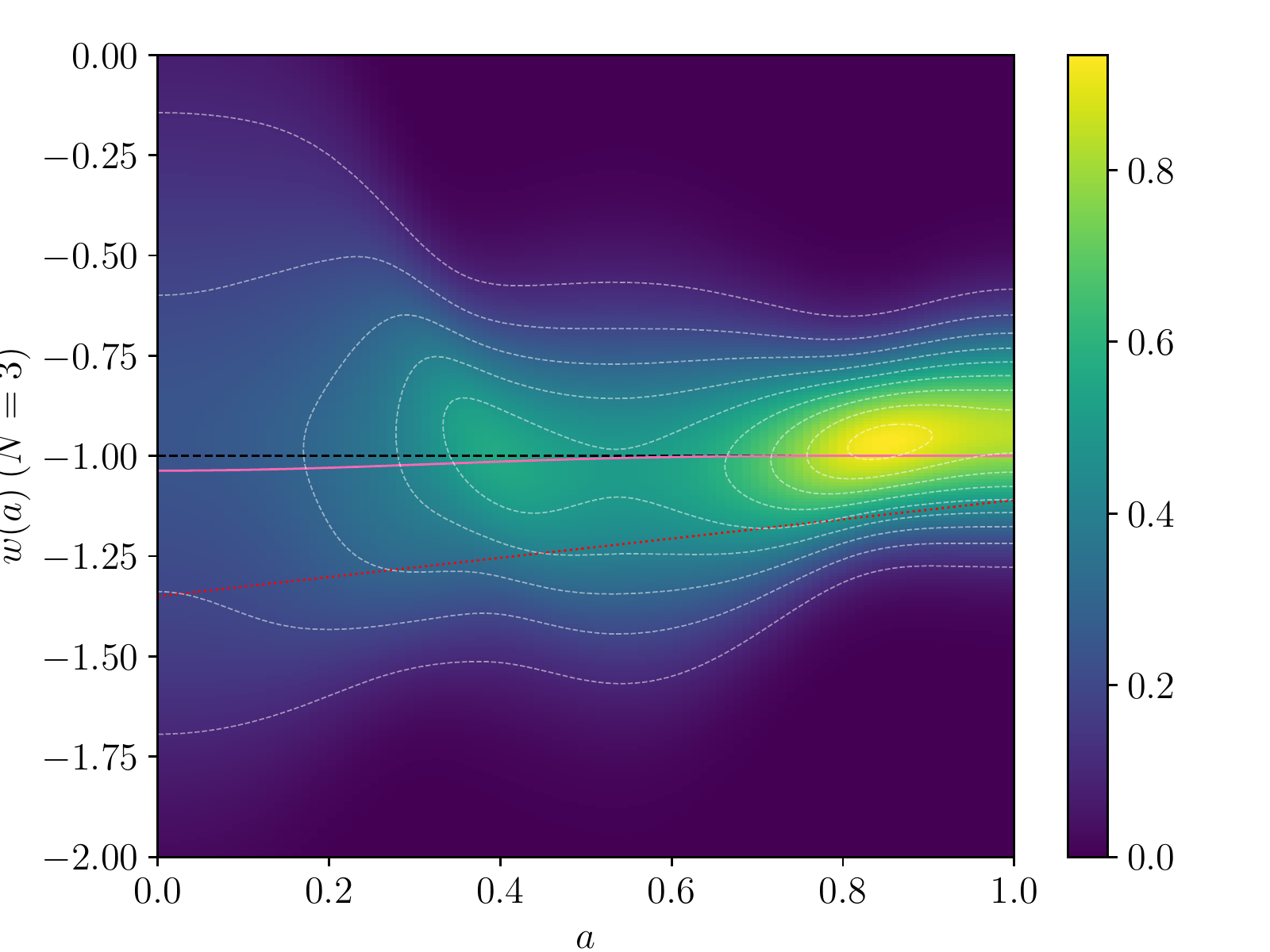}
    \caption{Posterior distributions for the dark energy equation of state for a non-Gaussian likelihood using 2 (left) and 3 (right) modes, respectively. The average curve is given in pink. The dotted red line displays the CPL model.}
    \label{curvature_1}
\end{figure}

In order to arrive at a statistical manifold with curvature, the distance modulus is expanded functionally to second order in the perturbations to the dark energy equation of state $\delta \w(a)$, 
\begin{align}
     y[a_i,\w] &\cong 
     y[a_i, \delta \w = 0 ] + \int\mathrm{d}a\: \frac{\delta y[a_i, \w]}{\delta \w(a)}\bigg|_{\w = -1} \delta\w(a)\: + \frac{1}{2}\int\mathrm{d}a\int\mathrm{d}a'\: \frac{\delta^2 y[a_i, \w]}{\delta \w(a)\delta \w(a')}\bigg|_{\w = -1} \delta\w(a)\delta \w(a')+  \mathcal{O}[(\delta \w)^3]\nonumber\\
     &\equiv y[a_i, \delta \w = 0] + \int\mathrm{d}a\: A^i(a) \:\delta\w(a)  + \frac{1}{2}\int\mathrm{d}a\:\int\mathrm{d}a'\: B^i(a,a') \: \delta \w(a) \delta \w(a') 
\end{align}
Using again the Fourier expansion (\ref{perturbation fourier}), the symmetric $B\indices{^i_{nn'}}$-coefficients are obtained,
\begin{align}
    B\indices{^i_{00}} &=\frac{3}{4} K(a_i) \bigg[\frac{\Omega_\mathrm{DE}}{\int_{a_i}^1 \mathrm{d}x'\: {x'}^{-2} (\Omega_m {x'}^{-3} + \Omega_\mathrm{DE})^{-1/2}
    }- 1\bigg]\int_{a_i}^1 \mathrm{d}x' \:\frac{\ln(x')^2}{{x'}^2 \:(\Omega_m {x'}^{-3} + \Omega_\mathrm{DE})^{3/2}} +\frac{9}{8} K(a_i)   \int_{a_i}^1 \mathrm{d}x'\: \frac{\ln(x')^2}{{x'}^2\: (\Omega_m{x'}^{-3} + \Omega_\mathrm{DE})^{5/2}}\\
    B\indices{^i_{0n}} &= \frac{3}{2}K(a_i)\frac{\Omega_\mathrm{DE}}{\int_{a_i}^1 \mathrm{d}x'\: {x'}^{-2} (\Omega_m {x'}^{-3} + \Omega_\mathrm{DE})^{-1/2}
    } \int_{a_i}^1 \mathrm{d}x' \: \frac{\ln(x')}{{x'}^2 \:(\Omega_m {x'}^{-3} + \Omega_\mathrm{DE})^{3/2}} 
    \int_{a_i}^1 \mathrm{d}z \: \frac{[\cosi(\pi n)- \cosi(\pi n z)]}{z^2\: (\Omega_m {z}^{-3} + \Omega_\mathrm{DE})^{3/2}}\nonumber\\
    &\quad-3 K(a_i)  \int_{a_i}^1 \mathrm{d}x' \: \frac{\ln (x')}{{x'}^2}  [\cosi(\pi n)- \cosi(\pi n x')] \bigg[(\Omega_m {x'}^{-3} + \Omega_\mathrm{DE})^{-3/2}-\frac{3}{2}(\Omega_m{x'}^{-3} + \Omega_\mathrm{DE})^{-5/2}\bigg]
    \\
    B\indices{^i_{nn'}} &= 3 K(a_i)  \frac{\Omega_\mathrm{DE}}{\int_{a_i}^1 \mathrm{d}x'\: {x'}^{-2} \:(\Omega_m {x'}^{-3} + \Omega_\mathrm{DE})^{-1/2}
    }  \int_{a_i}^1 \mathrm{d}z \: \frac{[\cosi(\pi n)- \cosi(\pi n z)]}{z^2 \:(\Omega_m z^{-3} + \Omega_\mathrm{DE})^{3/2}} \int_{a_i}^1 \mathrm{d}x' \: \frac{[\cosi(\pi n)-\cosi(\pi n x')]}{{x'}^2\: (\Omega_m {x'}^{ -3} + \Omega_\mathrm{DE})^{3/2}}\nonumber\\&\quad- 3 K(a_i) \int_{a_i}^1 \mathrm{d}x'\: \frac{1}{{x'}^2} [\cosi(\pi n)- \cosi(\pi n x')][\cosi(\pi n')- \cosi(\pi n' x')]\bigg[(\Omega_m {x'}^{-3} + \Omega_\mathrm{DE})^{-3/2}-\frac{3}{2}(\Omega_m{x'}^{-3} + \Omega_\mathrm{DE})^{-5/2}\bigg],
\end{align}
with $\mathrm{Ci}$ the cosine integral function, $\mathrm{Ci}(x) \equiv -\int_x^\infty \mathrm{d}\ln t \: \cos(t)$.
Together with the $A\indices{^i_n}$ derived before, these coefficients lead to the partition sum, in this case for a properly non-Gaussian case, and allows for MCMC sampling of the posterior distribution of $\w(a)$, as given in Fig.~\ref{curvature_1} for two and three Fouier-modes, respectively. As expected, the uncertainty is larger and the behaviour of $\w(a)$ at low $a$ is less constrained for three modes.

\begin{figure}
    \centering
    \includegraphics[width = 8cm]{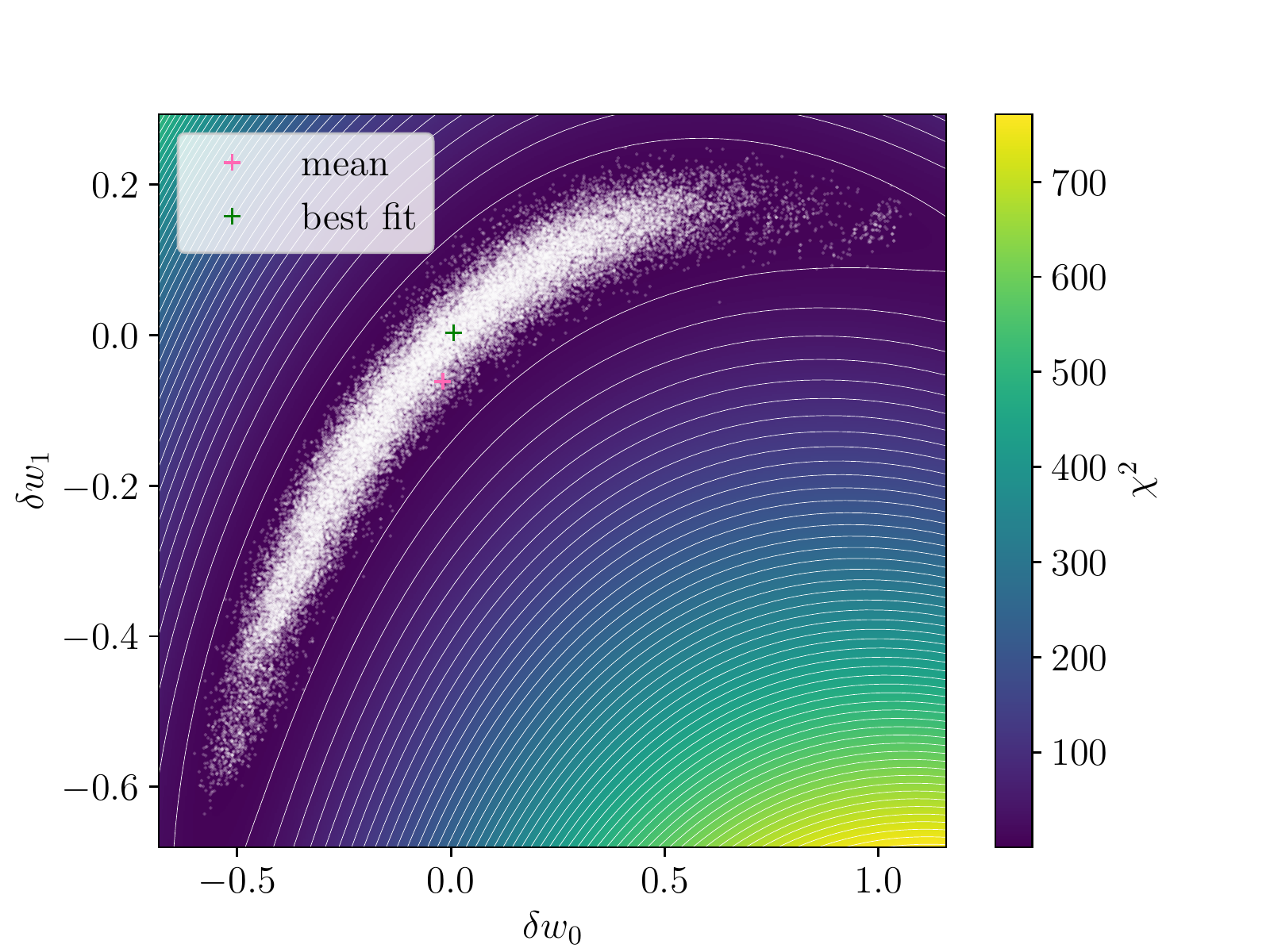}
    \includegraphics[width = 8cm]{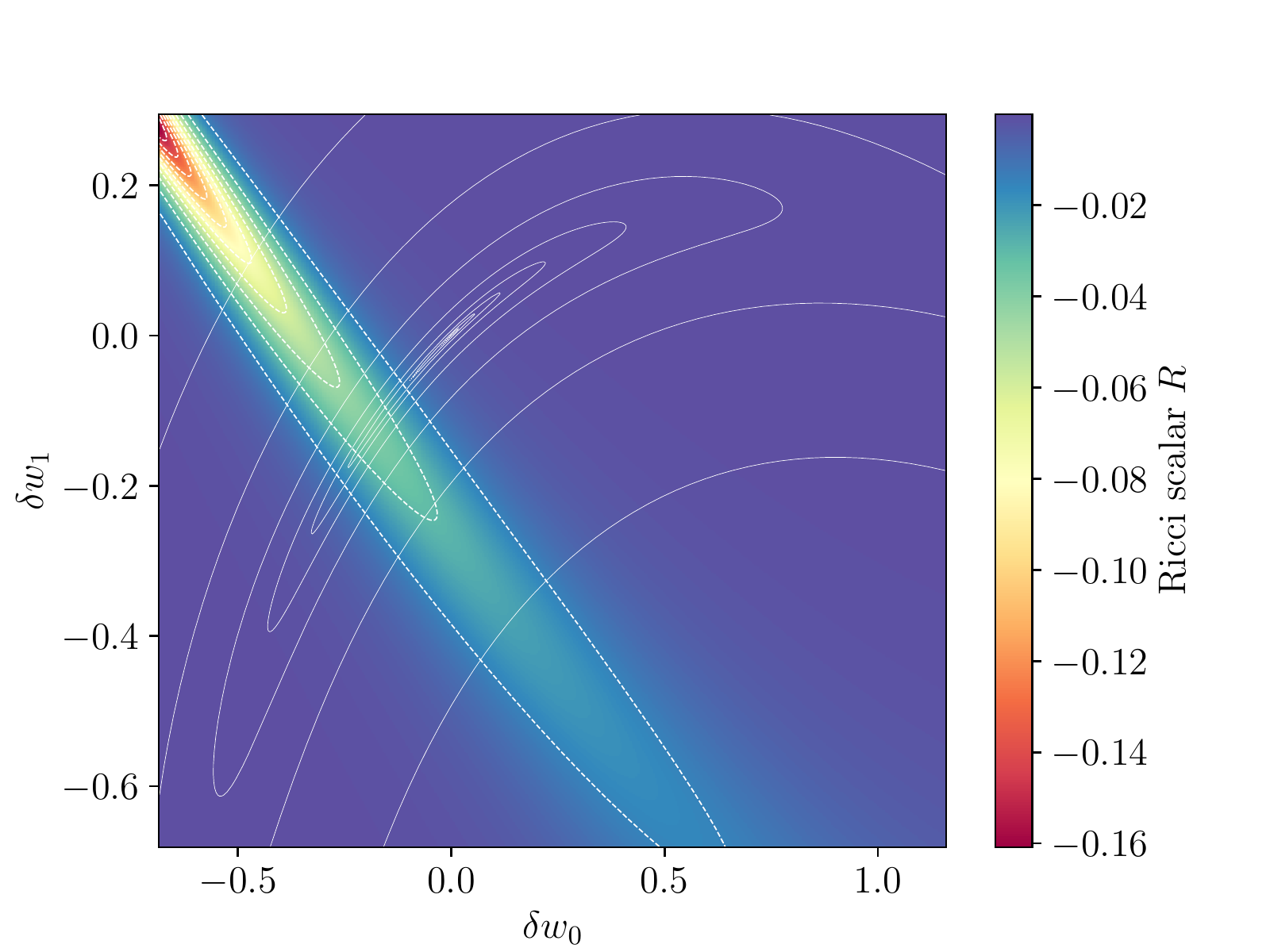}
    \caption{The residual $\chi^2$ together with the MCMC-generated samples with a choice of 2 modes $\delta w_0$ and $\delta w_1$ (left), and the residual $\chi^2$ superimposed with the Ricci scalar in the parameter space spanned by the two modes $\delta w_0$ and $\delta w_1$.}
    \label{ricci}
\end{figure}

Fig.~\ref{ricci} shows the residual $\chi^2$ in the parameter space of two Fourier-modes in the nonlinear model, and the deviation of the isoprobability contours from an elliptical shape are an indication of non-Gaussianity. Overlaid are MCMC-samples, from which the expectation value of the parameter pair is derived: As expected for a non-Gaussian likelihood, it differs from the best-fit value. As an outlook, we have computed the Ricci-scalar as derived from the Fisher-metric using a Levi-Civita-connection, and superimpose it onto the $\chi^2$ in the spirit of information geometry \citep{amari_information_2016}: We assert that there is a related geometry for functional manifolds. The non-Gaussianity becomes evident around the mean and best fit value, indicating extremal values of the posterior in this region of the parameter manifold.

\section{summary and discussion}\label{sect_summary}
The subject of our paper was the theory of Bayesian inference for functions, applied to the determination of the dark energy equation of state function $\w(a)$ from supernova data.

\begin{enumerate}
\item{For this purpose, we retrace the construction of a Fisher-metric from the Kullback-Leibler divergence $\Delta S$ between two likelihoods with infinitesimally differing parameter choices, and the subsequent construction of a partition function as a generalisation of the Bayesian evidence: For likelihoods that are conditional on a function $w(a)$ instead of a finite parameter set, the expansion of the Kullback-Leibler divergence yields a functional Fisher-matrix $F[w(a),w(a')]$. In complete analogy, the Bayesian evidence then becomes an integral over the set of possible functions and allows the construction of a partition $Z[\beta,J(a)]$, which effectively becomes a path integral.}
\item{The path integral has a closed solution for the Gaussian case. Functional differentiation of the partition then yield expectation value and variance of the inferred function, which for a Gaussian path integral come out as unbiased and in realisation of the variance given by the inverse Fisher-functional, as a reflection of the Cram{\'e}r-Rao inequality. The differentiation of $\ln Z/\beta$ with respect to $\beta$ yields a functional Shannon-entropy $S$ of the posterior distribution.}
\item{The requirement of models having to be differentiable in their parameters translates to functional differentiability in our case: We demonstrate this with the distance modulus in FLRW-spacetimes and construct posterior distributions of the equation of state function $w(a)$ as derived from a data set of supernova distance moduli}. While one naturally recovers Gaussian statistics for linear models, we expand the distance modulus in terms of a Volterra-series to second order in $w(a)$ to observe properly non-Gaussian statistics.
\item{Naturally, constraints are derived from a finite number of data points and increasing the model complexity results in larger errors and higher degrees of covariance between the model parameters, as expressed by the divergence of $\mathrm{det}F$. Introducing a parameterisation of the function $w(a)$ falls back onto the standard case of inference with finite parameter spaces. We showed by using Fourier-expansions and expansions into orthonormal Gegenbauer-polynomials that the shape of the inferred function does depend on the choice of parameterisation, that more complex models are less constrained, and that all models exhibit strong, parameterisation-dependent extrapolation errors towards redshifts where the data is sparse. For both Fourier-expansions and polynomials it is the case that the order imposes a variability scale.}
\item{Implicitly we assumed non-informative uniform priors on the equation of state function $w(a)$, but the definition of the functional partition $Z[\beta,J(a)]$ perfectly allows the inclusion of a prior distribution. \citet{crittenden_fables_2012} point out the importance of a prior choice in the inference of $w(a)$.}
\end{enumerate}

Finally, we would like to point out that other applications of the functional inference formalism might be reconstructions of the Hubble-function $H(a)$ and the growth function $D_+(a)$ without reference to a particular FLRW-model, or the inference of the CDM-spectrum $P(k)$ \citep{jasche_bayesian_2009, jasche_bayesian_2010, bohm_bayesian_2017}. A second case could be the free functions of the Horndeski-Lagrange density in modified gravity, which are not restricted by theory to obey a particular parameterisation \citep{bellini_constraints_2016}. One curious observation is the fact that Gaussian statistics is inextricably linked to linear models, Terms that introduce non-Gaussianity are at least at fourth order in $w(a)$, alluding to $\varphi^4$-interactions in quantum field theory. We intend to generalise our investigations to distributional Fisher-matrices and their corresponding partitions, for the inference of distributions. Applications might include source redshift distributions for e.g. gravitational lensing and their uncertainties. Here, positivity and normalisation need to be respected as additional constraints, as demonstrated by \citet{2023arXiv230104085R}, with the help of Lagrange-multipliers.

\section*{acknowledgements}
This work was supported by the Deutsche Forschungsgemeinschaft (DFG, German Research Foundation) under
Germany's Excellence Strategy EXC 2181/1 - 390900948 (the Heidelberg STRUCTURES Excellence Cluster).

\section*{data availability statement}
Our python-toolkit for deriving the Fisher-functionals for a data set of supernova distance moduli is available upon reasonable request.

\bibliographystyle{mnras}
\bibliography{references}

\appendix

\section{Details on the derivation of the functional Fisher-formalism}\label{appendix_functional_fisher}
Starting at a second-order expansion of the functional Kullback-Leibler divergence $\Delta S$, where the likelihood is conditional on a function $w(a)$ instead of a discrete parameter tuple $\theta^\mu$,
\begin{align}
    \Delta S \cong \int \mathrm{d}y \: \mathcal{L}(y|\w) \:&\left[ \int \mathrm{d}a\: \frac{(-1)}{\mathcal{L}(y|\w + \delta \w)} \frac{\delta \mathcal{L}(y|\w +\delta \w)}{\delta (\w(a)+\delta \w(a)} \Bigg|_{\delta \w = 0}  \delta \w(a)\right.\nonumber\\
    &+ \left.\frac{1}{2}\int\mathrm{d}a\int \mathrm{d}a' \: \frac{\delta \ln\mathcal{L}(y|\w+\delta\w)}{\delta(\w(a)+\delta \w(a))} \frac{\delta \ln \mathcal{L}(y|\w +\delta \w)}{\delta (\w(a')+\delta \w(a'))} \Bigg|_{\delta \w = 0}  \delta\w(a)\: \delta \w(a')\right. \nonumber
    \\ & \left.- \frac{1}{2}\int \mathrm{d} a \int \mathrm{d}a'\:\frac{1}{\mathcal{L}(y|\w +\delta \w)}\frac{\delta^2 \mathcal{L}(y|\w+\delta \w)}{\delta (\w(a)+\delta \w(a))\:\delta \w(a')+\delta \w(a'))}\Bigg|_{\delta \w =0}  \delta\w(a)\:\delta\w(a') \right].
\end{align}
one can use $\delta \w = 0$ and exchange functional derivatives with the integrations, which leads to 
\begin{align}
     \Delta S \cong \int \mathrm{d}y \: \mathcal{L}(y|\w) \:&\left[ \int \mathrm{d}a\: \frac{(-1)}{\mathcal{L}(y|\w)} \frac{\delta \mathcal{L}(y|\w)}{\delta \w(a)}   \delta \w(a)\right.+  \frac{1}{2}\int\mathrm{d}a\int \mathrm{d}a' \: \frac{\delta \ln\mathcal{L}(y|\w)}{\delta\w(a)} \frac{\delta \ln \mathcal{L}(y|\w)}{\delta \w(a')}  \delta\w(a) \:\delta \w(a')\nonumber
    \\ & \left.- \frac{1}{2}\int \mathrm{d}a \int \mathrm{d}a'\:\frac{1}{\mathcal{L}(y|\w)}\frac{\delta^2 \mathcal{L}(y|\w)}{\delta \w(a)\:\delta \w(a')}  \delta\w(a)\:\delta\w(a') \right]\nonumber\\
    = - \int \mathrm{d}a\: \delta \w(a)& \;\frac{\delta}{\delta \w(a)} \bigg(\int\mathrm{d}y\:  \mathcal{L}(y|\w)\bigg) + \frac{1}{2} \int\mathrm{d}a\int \mathrm{d}a' \: \int\mathrm{d}y \:\mathcal{L}(y|\w) \:\frac{\delta \ln \mathcal{L}(y|\w)}{\delta \w(a)} \frac{\delta \ln\mathcal{L}(y|\w)}{\delta \w(a')}  \delta \w(a)\: \delta \w(a')\nonumber\\
    &- \frac{1}{2} \int\mathrm{d}a \int \mathrm{d}a' \:\delta \w(a)\: \delta \w(a')\: \frac{\delta^2}{\delta\w(a) \:\delta \w(a')} \bigg(\int \mathrm{d}y\:\mathcal{L}(y|\w)\bigg).
\end{align}
Since the likelihood is normalized on the data space, only the second summand remains 
\begin{equation}
     \Delta S \cong \frac{1}{2} \int\mathrm{d}a\int \mathrm{d}a' \: \int\mathrm{d}y \:\mathcal{L}(y|\w) \:\frac{\delta \ln \mathcal{L}(y|\w)}{\delta \w(a)} \frac{\delta \ln\mathcal{L}(y|\w)}{\delta \w(a')}  \delta \w(a)\: \delta \w(a'),
\end{equation}
which defines the functional Fisher-metric.

\section{Fourier Coefficients}\label{Fourier Coefficients}
The partition function was written in terms of the Fourier expansions of the source terms $J(a)$ and the function $\w(a)$ to be inferred,
\begin{align}
       Z [\beta, J(a)] &= \int \mathrm{D}\w \: \exp\Big(-\frac{\beta}{2}\int\mathrm{d}a\int \mathrm{d}a'\: F_{aa'} \:\Big(\frac{\w_0}{2} + \sum_{n=1}^\infty \w_n \cos(\pi n a) \Big)\Big(\frac{\w_0}{2} + \sum_{n'=1}^\infty \w_{n'} \cos(\pi n' a')\Big)\Big)\nonumber \\
       &\hspace{1.15cm}\times \exp\Big(\int \mathrm{d}a \: \Big(\frac{J_0}{2}+\sum_{n=1}^\infty J_n \cos(\pi n a)\Big) \Big(\frac{\w_0}{2} + \sum_{n'=1}^\infty \w_{n'} \cos(\pi n' a)\Big)\Big).\label{fourier Z coeff}
\end{align}
For the term containing the sources $J(a)$, one finds,
\begin{align}
    \quad \int \mathrm{d}a \: \bigg(\frac{J_0}{2}+\sum_{n=1}^\infty J_n \cos(2\pi n a)\bigg) \bigg(\frac{\w_0}{2} + \sum_{n'=1}^\infty \w_{n'} \cos(\pi n' a)\bigg) &= \frac{J_0 \w_0}{4} + \sum_{n=1}^\infty \sum_{n'=1}^\infty J_{n}\w_{n'} \:\int_0^1 \mathrm{d}a \: \cos(\pi n a) \cos(\pi n' a) =\frac{J_0 \w_0}{4} + \frac{1}{2}\sum_{n=1}^\infty J_{n}\w_{n},
\end{align}
because of the orthogonality of the Fourier-basis
\begin{equation}
    \int_0^1 \mathrm{d}a\: \cos(\pi n a) \cos(\pi n' a) = \begin{cases}  0 \quad \text{for}\quad n = n'\\    \frac{1}{2}\quad \text{for} \quad n\neq n' 
    \end{cases} = \frac{1}{2}\:\delta_{nn'}.
\end{equation}
Therefore, the Fourier expansion of the Fisher matrix components leads to
\begin{align}
    F_{aa'} &= 
    A_i(a) A^i(a') = 
    \frac{\Tilde{A}\indices{^i_0}\Tilde{A}\indices{_{i0}}}{4} + 
    \frac{\Tilde{A}\indices{^i_0}}{2}\sum_{n=1}^\infty \Tilde{A}\indices{_{in}} \cos(\pi n a) +
    \frac{\Tilde{A}\indices{^i_0}}{2}\sum_{n'=1}^\infty \Tilde{A}\indices{_{in'}} \cos(\pi n' a') + 
    \sum_{n=1}^\infty\sum_{n'=1}^\infty  \Tilde{A}\indices{^i_n} \Tilde{A}\indices{_{in'}} \cos(\pi n a)\cos(\pi n'a'), \label{fisher fourier}
\end{align}
which is inserted into \eqref{fourier Z coeff}. For simplicity, we work with a diagonal unit data covariance, which can be always reached by a linear transform.

In order to return to the finite case the discrete Fisher metric $F_{nn'}$ needs to be identified in terms of $\Tilde{A}\indices{^i_n}, \Tilde{A}\indices{^i_{n'}}$. The Fisher metric for the Gaussian-case with uncorrelated data is derived in the usual way via the residuals $\chi^2$:
\begin{align}
    \chi^2 &= 
    \bigg(y_i - \int_0^1 \mathrm{d}a \: \Tilde{A}\indices{_i}(a) \w(a)\bigg) 
    \bigg(y^i - \int_0^1 \mathrm{d}a'\: \Tilde{A}\indices{^i}(a') \w(a')\bigg)\nonumber \\&=  
    \bigg(y_i - \int_0^1 \mathrm{d}a \: \bigg(\frac{\Tilde{A}\indices{_{i0}}}{2} + \sum_{n=1}^\infty \: \Tilde{A}\indices{_{in}} \cos(\pi n a)\bigg)\bigg(\w_0 + \sum_{n' =1}^\infty \w_{n'} \cos(\pi n' a)\bigg)\bigg)
    \bigg(y^i - \int_0^1 \mathrm{d}a' \: \bigg(\frac{\Tilde{A}\indices{^i_0}}{2} +\sum_{n =1}^\infty \: \Tilde{A}\indices{^i_m} \cos(\pi m a')\bigg)\bigg(\w_0 + \sum_{m' = 1}^\infty \w_{m'} \cos(\pi m' a')\bigg)\bigg)\nonumber
    \\
        &= 
    \bigg(
    y_iy^i - 
    y_i \Tilde{A}\indices{^i_0} \frac{\w_0}{2} -
    y_i \sum_{n = 1}^\infty \Tilde{A}\indices{^i_n} \w_n + 
    \frac{\Tilde{A}\indices{^i_0}\Tilde{A}\indices{_{i0}}}{4} \frac{\w_0^2}{4} + 
    \frac{\Tilde{A}\indices{^i_0}}{2} \frac{\w_0}{2} \sum_{n = 1}^\infty  \Tilde{A}\indices{^i_n} \w_n  + \frac{1}{4} \sum_{n = 1}^\infty \sum_{n' = 1}^\infty \Tilde{A}\indices{^i_n} \Tilde{A}\indices{_{in'}}\w_n \w_{n'}
    \bigg)\nonumber
    \\
    &= - 2 Q_0 \frac{\w^0}{2} - 2 \sum_{n = 1}^\infty Q_n \w^n + F_{00} \frac{\w_0^2}{4} + \sum_{n = 1}^\infty F_{0n}\frac{\w^0}{2} \w_n + \sum_{n = 1}^\infty \sum_{n' = 1}^\infty F_{nn'}\w^n \w^{n'}. 
\end{align}
up to an additive constant, and again for a unit data covariance for simplicity. Hence, 
\begin{equation}
    F_{00} = \frac{\Tilde{A}\indices{_{i0}} \Tilde{A}\indices{^i_0}}{4}, 
    \quad 
    F_{0n} = \frac{\Tilde{A}\indices{_{i0}} \Tilde{A}\indices{^i_n}}{2},
    \quad 
    F_{nn'} = \frac{\Tilde{A}\indices{_{in}} \Tilde{A}\indices{^i_{n'}}}{4}.
    \label{Fisher discrete}
\end{equation}
Writing out all the products in \eqref{fourier Z coeff} and performing the integrations, the different terms can then be identified as the discrete Fisher metric components. The summands of our expression can be identified with the Fourier coefficients of the Fisher metric components \eqref{Fisher discrete}, e.g.  
\begin{align}
    \int_0^1\mathrm{d}a\int_0^1 \mathrm{d}a'\: F_{aa'} \:\bigg(\frac{\w_0^2}{4}\bigg) &= \int_0^1 \mathrm{d}a \int_0^1\mathrm{d}a'\: \frac{1}{\sigma_i^2} \bigg( \frac{\Tilde{A}^i_0}{2}+\sum_{n=1}^\infty \Tilde{A}^i_n \cos(\pi n a)\bigg)\; \bigg(\frac{\Tilde{A}^i_0}{2}+\sum_{n'=1}^\infty \:  \Tilde{A}^i_{n'} \cos(\pi n' a') \bigg) \:\bigg(\frac{\w_0^2}{4}\bigg)\nonumber \\&=  \int_0^1 \mathrm{d}a\int_0^1 \mathrm{d}a'\: \frac{1}{\sigma_i^2} \bigg(\frac{\Tilde{A}^i_0}{2}\bigg)^2 \bigg(\frac{\w_0}{2}\bigg)^2 = F_{00}\: \bigg(\frac{\w_0}{2}\bigg)^2.
\end{align}
In conclusion, the partition function for the discrete case (with data mean set to $y_i = 0$ for simplicity) is given by
\begin{equation}
    Z [\beta, J_n] = 
    \int \mathrm{d}^n\w \: \exp\bigg(-\frac{\beta}{2} \bigg(F_{00}\frac{\w_0^2}{2} + \sum_{n = 1}^\infty F_{0n} \frac{\w_0}{2}\w_n + \sum_{n = 1}^\infty \sum_{n' = 1}^\infty F_{nn'}\w_n \w_{n'} + \frac{1}{2} J_0 \frac{\w_0}{2} + \frac{1}{2} \sum_{n = 1}^\infty J_n \w_n \bigg)\bigg),
\end{equation}
in accordance with our expectations. Please note that because of how this case was constructed, the $\Tilde{A}\indices{^i_n}$-coefficients do not directly correspond to the $A\indices{^i_n}$ from the quadratic expansion. The coefficients are related via $A\indices{^i_0} = \frac{1}{2} \Tilde{A}\indices{^i_0}, \; A\indices{^i_n} = \frac{1}{2} \Tilde{A}\indices{^i_n}$. These two constructions correspond to two different ways in which the finite case was approached. However, both ways finally turn out to coincide in terms of the Fisher-metric and the partition function.

\section{Functional Derivatives}\label{appendix_functional_derivatives}
The first functional derivative of the distance modulus with respect to the dark energy equation of state reads,
\begin{equation}
    \frac{\delta y[a_i,\w]}{\delta \w(a)} = 
    - K(a_i) \int_{a_i}^1\mathrm{d}x' \:\frac{\exp\Big(-3 \int_1^{x'}(1+\w(x))/x\Big)}{{x'}^2\: (\Omega_m {x'}^{-3}+\Omega_\mathrm{DE})^{3/2}}  \frac{\textbf{1}_{[x',1]}(a)}{a},
\end{equation}
with $a_i$ being the scale factor in question and $\w(y)$ the DE eos function. As previously, the abbreviation
\begin{equation}
    K(a_i)\equiv 
    \frac{15\:\Omega_\mathrm{DE}}{2\ln(10)}\times\bigg(\int_{a_i}^1 \mathrm{d}x' \frac{1}{{x'}^2 \sqrt{\Omega_m {x'}^{-3}+\Omega_\mathrm{DE}}}\bigg)^{-1},
\end{equation}
is used, which is a function of the scale factor, but does not depend functionally on $\w(a)$. The second functional derivative can be written as 
\begin{align}
    \frac{\delta^2 y[a_i, \w]}{\delta \w(a) \delta \w(a')} &= 
    \frac{3}{2} \frac{K(a_i) \Omega_\mathrm{DE}}{ \bigg(\int_{a_i}^1 \mathrm{d}x'\: {x'}^{-2}\: (\Omega_m {x'}^{-3} + \Omega_\mathrm{DE})^{-1/2}\bigg)} \int_{a_i}^1 \mathrm{d}x'\:\frac{\exp\Big(-3 \int_1^{x'}(1+\w(x))/x\Big)}{{x'}^2\: (\Omega_m {x'}^{-3}+\Omega_\mathrm{DE})^{3/2}}\frac{\textbf{1}_{[x',1]}(a')}{a'}  \nonumber\\&\hspace{+7cm}
    \times \int_{a_i}^1\mathrm{d}z\: \frac{\exp\Big(-3 \int_1^{x'}(1+\w(x))/x\Big)}{{z}^2\: (\Omega_m {z}^{-3}+\Omega_\mathrm{DE})^{3/2}} \frac{\textbf{1}_{[z,1]}(a)}{a} \nonumber \\
    &- 3 K(a_i) \: \int_{a_i}^1 \mathrm{d}x' \: \exp\bigg(-3\int_1^{x'} \mathrm{d}x\:\frac{1+\w(x)}{x} \bigg)\:\bigg(\frac{\textbf{1}_{[x',1]}(a)\: \textbf{1}_{[x',1]}(a')}{aa'}\bigg)\: \Bigg[{x'}^{-2} (\Omega_m {x'}^{-3}+\Omega_\mathrm{DE})^{-3/2} - \frac{3}{2}{x'}^{-2} (\Omega_m {x'}^{-3}+\Omega_\mathrm{DE})^{-5/2}\Bigg].
\end{align}
Here, the product of two Heaviside functions was rewritten using the indicator function $\textbf{1}_A$ for  $A\subset \mathbb{R}$, e.g.
\begin{equation}
    \theta(a-x') \: \theta(1-a) = \textbf{1}_{[x',1]}(a) = 
    \begin{cases}
        0 \;\text{for}\; a \in [x',1]\\
        1 \;\text{for}\; a \notin [x',1]
    \end{cases}.
\end{equation}

\label{lastpage}
\end{document}